\documentclass[12pt,preprint]{aastex}
\usepackage{ulem} 

\slugcomment{To appear in AJ}

\shorttitle{Spectrscopic Monitoring of Class I YSOs}
\shortauthors{Connelley et al. 2011}

\begin{document}


\title{Near-IR Spectroscopic Monitoring of Class I Protostars: Variability of Accretion and Wind Indicators}


\author{Michael S. Connelley\altaffilmark{1}}
\affil{University of Hawaii, Institute for Astronomy, 640 N. Aohoku Pl., Hilo, HI 96720}

\author{Thomas P. Greene}
\affil{NASA Ames Research Center, M.S. 245-6, Moffett Field, CA., 94035}


\altaffiltext{1}{Support Astronomer at the Infrared Telescope Facility, which is operated by the University of Hawaii under Cooperative Agreement no. NNX08AE38A with the National Aeronautics and Space Administration, Science Mission Directorate, Planetary Astronomy Program."}

\begin{abstract}

   We present the results of a program that monitored the near-IR spectroscopic variability of a sample of 19 embedded protostars.   Spectra were taken on time intervals from 2 days to 3 years, over a wavelength range from 0.85~$\mu$m to 2.45~$\mu$m, for 4-9 epochs of observations per target.  We found that the spectra of \textit{all} targets are variable, and that \textit{every} emission feature observed is also variable (although not for all targets).  With one exception, there were no drastic changes in the continua of the spectra, nor did any line completely disappear, nor did any line appear that was not previously apparent.  This analysis focuses on understanding the connection between accretion (traced by H Br $\gamma$ and CO) and the wind (traced by He I, [FeII], and sometimes H$_2$).  For both accretion and wind tracers, the median variability was constant versus time interval between observations, however the maximum variability that we observed increased with time interval between observations.  Extinction is observed to vary within the minimum sampling time of 2 days, suggesting extinguishing material within a few stellar radii at high disk latitudes.  The variability of [FeII] and H$_2$ were correlated for most (but not all) of the 7 YSOs showing both features, and the amplitude of the variability depends on the veiling.  Although the occurrence of CO and Br $\gamma$ emission are connected, their variability is uncorrelated, suggesting that these emissions originate in separate regions near the protostar (e.g., disk and wind).  The variability of Br $\gamma$ and wind tracers were found to be positively correlated, negatively correlated, or uncorrelated, depending on the target.  The variability of Br $\gamma$, [FeII], and H$_2$ always lies on a plane, although the orientation of the plane in 3D depends on the target.  While we do not understand all interactions behind the variability that we observed, we have shown that spectroscopic variability is a powerful tool towards understanding the star formation process.  

\end{abstract}



\keywords{stars:formation, stars: pre-main sequence, infrared: stars, techniques: spectroscopic}


\section{Introduction}



   Much of what we have learned about young stars has come through studies of variability.  T Tauri stars were first identified as a class of unusual variable stars (Joy 1945) before it was proposed that they were young stars undergoing gravitational collapse (Herbig 2002).  More recent studies have further explored the photometric variability of young stars.  Carpenter et al. (2001) found that among the variable YSOs they observed in the Orion A cloud, the variability of 56\% to 77\% stars could be attributed to star spots, at least 23\% of variability could be attributed to variable extinction, and about1\% could be attributed to changes in the mass accretion rate.  Grankin et al. (2007) showed that the variability of T Tauri stars is stable on long time scales, reflecting shorter term variability due to stellar rotation and star spots.  In contrast to that result, Manset et al. (2009) found that the variability of RY Lupi is due in part to an asymmetric disk.  A 20 year optical monitoring program by Artemenko et al. (2010) found that stars with higher bolometric luminosity had slower variability, and that that variability time scale is a fourth of the Keplerian period at the dust-sublimation radius.  Near-IR photometric monitoring of YSOs by Alves de Oliveira et al. (2008) found that the variability of their sample targets in $\rho$ Oph can be attributed to stellar rotation and star spots, as well as changes in extinction or changes in the accretion disk.  Using archival near-IR data, Scholz (2012) found that about half of YSOs are variable, very few have variability greater than 0.5 magnitudes, the amplitude being the largest in NGC~1333 which has the youngest sample studied.  

   There have been several studies of YSO variability in the mid-IR.  Mid-IR variability of a transitional disk YSO (Muzerolle et al. 2009) suggests changes in the inner disk height or perturbations to the inner disk.  Changes in optical/near-IR (Eiroa et al. 2002) and mid-IR (Barsony et al. 2005) have been interpreted as evidence for variable veiling among T Tauri stars.  Variability of the strength and profile of the 10~$\mu$m silicate feature for two T Tauri stars (Bary et al. 2009) lead those authors to suggest changes in optical depth to explain the variability.   These studies demonstrate that YSO variability has been observed over a wide wavelength range, and can probe a diverse range of origins.  A study of the mid-IR variability across the Orion star forming regions by Megeath et al. (2013) found that about half of the YSOs were variable, with light curve amplitudes of about 0.2 magnitudes.  They note that protostars are more likely to be variable, and tend to have larger light curve amplitudes.  The Spitzer YSOVAR study is showing that Class Is have photometric variability similar to CTTSs, suggesting that they have similar accretion and extinction processes.  Morales-Calder\'{o}n (2011) found that Class I and II YSOs show variability from ~0.3 to 1 magnitude at 3.6~$\mu$m and 4.5~$\mu$m within 40 days.  For some targets, the variability is periodic, with periods ranging from 1 to 25 days.  Contemporaneous spectroscopy (Faesi et al. 2012) of 5 targets from the YSOVAR sample found that the He~I emission was variable over a two month baseline, but mid-IR variability and spectroscopic variability appeared uncorrelated.  


  Observations of spectral (emission) lines have revealed much about how TTSs accrete and eject matter via mass flows (H Br $\gamma$, e.g. Muzerolle et al. 1998) and winds ([FeII], H$_2$, e.g. Zinnecker et al. (1998) and Beck et al. 2008). This has improved our understanding of the locations, dynamics, and masses of disks as well as our understanding of the accretion funnel flows and winds of young stars.  The first study to note the variability of emission lines among CTTS was Joy (1945), who observed that the emission line flux was positively correlated with the photometry of the T Tauri star.  Nguyen (2009) observed that the mass accretion rate of a sample of 40 T Tauri stars typically doesn't change by more than 0.5 dex on time scales of days to months, and that the maximum extent of variability is reached after a few days.  Another example from time resolved spectroscopy is the finding that the CO temperature of V1647 Ori varied with the accretion rate, and that Br $\gamma$ flux and FWHM varied with the brightness of the YSO (Brittain et al. 2010).  Biscaya et al. (1997) found significant variability in the CO emission among a majority of their sample of YSOs on a time scale as short as a few days, with the variability of DG Tau possibly being periodic.  Only one object with CO absorption (FU Ori star V1097 Cyg) showed variability.  In addition to the line strength, observers have found that line profiles are also variable.  For example, Choudhury et al. (2011) found that the H~$\alpha$ line profile varied on time scales from a month to less than 1 hour.  

   The above examples illustrate that YSOs (mostly T Tauri stars) are variable both in their photometry and spectroscopic properties, from visible light to the mid-IR.  To explore what variability can teach us about the embedded protostellar phase of star formation, we have conducted a near-IR spectroscopic monitoring program of embedded protostellar YSOs.  The goal was to explore the mechanisms and time scales of their mass accretion and wind, using diagnostics developed for the somewhat older CTTSs that are accreting matter less actively.  In Section 2, we describe our sample selection and observations.  Section 3 details our data analysis methods for both photometry and spectroscopy.  In Section 4, we discuss how the accretion and wind tracers vary, both independently and in correlation with respect to each other.  We finish with notes of interest on individual targets.\\



\section{Sample and Observations}
   We obtained multi-epoch observations of 19 Class I YSOs selected from the sample of \citet{Con2010}, which were selected from the whole sky to have increasing flux density in Janskys with wavelength through the IRAS bands \footnote{Since spectral index is defined as d(log$\lambda$F$_{\lambda}$)/d(log$\lambda$), it is possible for a source to have increasing flux density with wavelength yet still have a negative spectral index}, to be in regions of nearby high optical extinction, and in most cases to have a red (H$-$K$\gtrsim$1) near-IR counterpart observed by 2MASS.  Since our observations were conducted at the NASA IRTF, the sample targets are within the declination limits of the IRTF ($\delta < +70$) and rise above 2 airmasses from Mauna Kea ($\delta > -40$).  From the sample of \citet{Con2010}, we selected 19 targets for this spectroscopic monitoring program.  Since we did not know how many targets would be variable, we chose 19 targets as a compromise between have a large sample versus the number of observations per target.  As this program was largely exploratory, we selected the sample to have a wide range of spectroscopic features.  Such features include high and low veiling \footnote{We defined the veiling to be 'high' when no photospheric absorption lines are apparent in the spectrum.  When we can see photospheric absorption lines in the spectrum, then we considered the veiling to be 'low'.}, emission lines tracing accretion and winds, FU Ori-like features.   Of the 19 targets, 12 had high veiling, 4 have low veiling, 15 show Br $\gamma$ emission, 10 have H$_{2}$ or [FeII] emission, 10 have CO emission, and 3 have FU Ori-like spectra.  The observed sample is listed in Table 1, and the observing log is presented in Table 2.  


  Each object was observed with the 3.0~m IRTF on Mauna Kea, Hawaii with SpeX \citep{Ray2003} in the short cross-dispersed mode, which covers 0.8~\micron~ to 2.45~\micron~ in each exposure.  There is a gap in wavelength coverage between 1.82~\micron~ and 1.88~\micron, corresponding to a wavelength range where the atmosphere is relatively opaque.  We used the 0\farcs5 wide slit, which gives a resolution of R=1200.  The star was nodded along the slit, with two exposures taken at each nod position.  The total exposure time was driven by our goal to get a S/N$\approx$300 spectrum for each target for each epoch.  The individual exposure times were limited to two minutes to ensure that the telluric emission lines would cancel when consecutive images taken at alternate nod positions were differenced.  An A0 telluric standard star was observed after at least every other protostellar target for telluric correction, usually within 0.1 airmasses of the target.  An argon lamp was observed for wavelength calibration and a quartz lamp for flat fielding.  An arc/flat calibration set was observed for each target/standard pair.  

   The observations for this study were taken over the course of over 4 years.  Most data were taken in 2009 for this monitoring program, and other data taken for our spectroscopic survey \citep{Con2010} were used to supplement the data set.  The median number of spectral epochs observed per target is 7, the minimum number of epochs is 4, and the maximum number of epochs is 9.  Since we did not know the time scale of the variability of Class I protostars, we requested a cadence of roughly 2 weeks to allow for time the spectra of the targets to change but to also get several observations within a semester.  Ultimately, the interval between observations ranged from less than a day to several years.  

   When observing conditions were photometric, we took K-band images using the SpeX guider camera to flux calibrate the spectra.  For this imaging data, the telescope was dithered using a 7 point 7\arcsec radius pattern to allow us to make "sky flats'' from the data.  A dark frame was made by averaging together 10 individual dark frames of the same exposure time as the science data.  This dark was then subtracted from each target frame.  Sky flats were made by scaling each dark subtracted image to have the same median value, then averaging them together using a min-max rejection.  The resulting sky flat was then normalized using the median value of the pixel counts.  Each dark subtracted (but not scaled) target frame was divided by this normalized sky flat.  The median sky value for each frame was subtracted from each frame to set the average background counts in each frame to 0 to account for changes in the brightness of the sky.  The images were then aligned and averaged together using an average sigma clipping rejection.  Standard stars that have been observed by UKIRT through the MKO filter set (Simons \& Tokunaga 2002, Tokunaga \& Simons 2002) were selected from the UKIRT faint standard star list, and were observed for photometric calibration.

\section{Analysis}
   We now discuss our selection of emission lines than we monitored, how we measured them, and our photometric calibration.  We measured the correlation coefficient for the variability of various pairs of lines to determine what variability was correlated.  We developed new metrics to help disentangle truly correlated variability from variability of lines that appears correlated but is not.  Changes in the veiling would cause a false correlation when EW is used as the metric of line strength, whereas changes in extinction would cause a false correlation when line flux is used as the metric of line strength.

\subsection{Spectroscopy}

   The SpeX data were flat fielded, extracted, and wavelength calibrated using \emph{Spextool} \citep{Cus2004}.  After extraction and wavelength calibration, the individual extracted spectra were coadded with \emph{xcombspec}.  \emph{Xtellcor} was then used to construct a telluric correction model using the observed A0 standard, after which the observed spectrum of the target was divided by the telluric model.  Finally, \emph{xmergeorders} was used to combine the spectra in the separate orders into one continuous spectrum.  These are all IDL routines written by \citet{Cus2004} to completely reduce SpeX data.  Spectral line flux, equivalent width (EW), and FWHM were measured using the SPLOT routine in IRAF.  In the case of the He I lines, which can show both emission and absorption components, the EW measurement includes both the emission and absorption.


   The near-IR spectra of Class I YSOs have a wide variety of emission lines, as described in \citet{Con2010}.  The excitations and strengths of many emission lines are not well understood, and their analysis is beyond the scope of this work.  However, several near-IR emission lines are known to be related to accretion or wind/outflow activity, and we chose several of those to understand their variability and relationships.  HI Br $\gamma$ emission has been correlated with mass accretion in TTSs \citet{Muz1998}, and we investigate its correlations with CO, H$_{2}$, and [Fe II] in this study.  We also analyzed CO emission as an accretion tracer as it likely originates from a hot circumstellar disk photosphere \citep{Cal1991}.  To trace the wind, we chose to analyze H$_2$ v=1-0 S(1) (e.g. Davis et al. 2001, Beck 2008) and [FeII] (e.g. Bally et al. 2001).  Analysis of the [FeII] lines at 1.256~$\mu$m and 1.644~$\mu$m is also useful to determine extinction since both lines have the same upper level.  Finally, we chose to analyze the He I 1.083~$\mu$m to probe the stellar wind \citep{Edw2006}.

   We used equivalent width (EW) measurements for further analysis of the relative variability of spectral features.  EW is easily measured and is insensitive to the spectral resolution.  Unlike line flux, EW is not affected by non-photometric observing conditions nor changes in throughput of the instrument (such as changes in seeing affecting the slit loss).  We define relative variability as the EW of the newer observation divided by the EW of the older observation.  We considered the relative variability between all observation pairs, such that 7 observations of a target yield 21 observation pairs.  We note that changes in the continuum level may affect the measured EW.  However, only one target (IRAS 18247$-$0212) showed a systematic trend in its K-band photometry.  Also, changes in the continuum level would not affect the measured change in EW of one line relative to another line of a similar wavelength.  



   To understand how variable different lines were on various time intervals, we plotted the absolute value of the relative variability of an observation pair on a linear scale versus the log of the time interval between the observation pair.  We present plots for our two accretion tracers (Br $\gamma$ and CO, Figures 1 and 2), our two wind tracers ([FeII] and H$_{2}$, Figure 3), as well as for He I emission (Figure 4).  The variability seen at a given time interval is the absolute value of the ratio of the EWs observed at the two time intervals, and thus reflects the total variability over that time interval.  The plots include the median variability, taken over each 0.5 dex of time interval.  


   An important goal of this project is to understand how the variability of different lines are related.  To this end, we plotted the relative change of one line versus another.  Br $\gamma$ and CO were both chosen as mass accretion tracers.  \citet{Con2010} showed that the presence Br $\gamma$ and CO in emission is correlated; CO is only in emission when the veiling is high and the spectrum shows Br $\gamma$ emission.  Thus, we analyzed the relative change of these two emission features to see if the variability of these tracers are also correlated (Figure 5), and thus probe the connection between the disk surface temperature (traced by CO) and the rate of accretion onto the star (traced by Br $\gamma$).  We plotted the relative change of H$_2$ vs. [FeII] since they are our two wind tracers (see Figure 6).  \citet{Beck2008} showed that like [FeII], H$_2$ is also often shock excited.  As such, we expect that the variability of these two lines would be highly correlated.  Finally, winds are most likely driven by mass accretion.  We thus plotted Br $\gamma$ vs. H$_2$ and Br $\gamma$ vs. [FeII] to see how well the variability of the accretion and the wind are correlated (see Figures 7 and 8).  These figures will be discussed in detail below.

    Each data point in these plots is the relative change in one line versus the relative change in the other for a given observation pair for one target.  Color is used to designate data from different targets.  Having plotted the relative change for pairs of lines, we then calculated the correlation coefficient, as well as the slope and slope uncertainty in the linear regression line for each line pair for each target.  We used these results to test if there the likelihood of a real correlation between the two lines exceeds 95\% confidence.   For this test, we used the correlation coefficient (r) and the number of data points (N) to calculate:
\begin{displaymath}
t = \frac{r\sqrt{N-2}}{\sqrt{1-r^{2}}}
\end{displaymath}
   The calculated t value is then compared to the values for a two-tailed Student's t test with $\nu$=N$-$2 degrees of freedom to determine if the likelihood of a real correlation between the two lines exceeds 95\%.  These results are summarized in Table 3 and 4. 



   Five targets showed both important mass accretion (Br $\gamma$) and wind ([FeII] and H$_2$) tracers.  To understand how these accretion and wind tracers are related, we plotted the relative change of Br $\gamma$ and [FeII] and H$_2$ in 3D for each object.  We used the planefit.pro routine from the IDL astronomy library at NASA GSFC to find the best-fit plane to the data, and present the coefficients of the fits in Table 5. 

\subsection{Metric Testing}

   A key goal of this project is to determine which wind and/or accretion tracers show correlated variability.  EW is a useful measure of line strength that is unaffected by non-photometric weather and variable slit losses.  However, the measured EW is affected by changes in the veiling flux from the YSO's circumstellar disk.  The variability of the veiling can create a false correlation in the variability of a pair of lines when EW alone is used to measure the line strength.  Line flux can also be used to measure line strength, despite being affected by non-photometric weather and variable slit losses.  However, variability of the extinction can create a false correlation in the variability of a pair of lines when line flux alone is used to measure the line strength.  We are only able to estimate the extinction in 7/19 of our targets since we were not able to observe both the 1.256~$\mu$m  and 1.644~$\mu$m [FeII] emission lines from 12 targets.  The goal of this analysis is to develop a metric using both EW and line flux to discriminate between truly correlated variability vs. apparent variability caused by changes in flux and extinction.  In developing this metric, we assume that the variability of the extinction and veiling are uncorrelated.

   We needed the ability to test whether a metric could discriminate between correlated variability between two lines and 'noise' created by variable veiling and extinction, and to determine the false positive rate for a metric (i.e. how often the metric could determine that the lines are correlated when in they are not).  For this purpose, we wrote a data simulator that generated two synthetic light curves; each representing the light curve of a spectral lines.  Each simulation generates a light curve with 10 data points, and $10^4$ light curves per simulation.  The EW and flux of the two lines, as well as the veiling and extinction, vary randomly, each with its own random variable.  The the purpose of our metric testing, the veiling was uniform for both lines (i.e. no wavelength dependence), whereas real veiling is wavelength dependent.  The amount of veiling is not important whereas the change in veiling is important.  We assume that the change in veiling is wavelength independent, as we do not have enough parameters to constrain a wavelength dependent veiling change.  We can independently change the amplitude of variability of the two lines, as well as the variability of the veiling and extinction.  To switch from uncorrelated to correlated variability, we change both of the two spectral lines by the same random variable instead of each line having its own.  The output of the simulation can be varied to suit the metric being tested, and to numerically determine the confidence intervals to which the null result (no correlation) can be rejected.


   Based on the results from the metric testing with the data simulator, we chose the following metric using both line EW and line flux.  We use standard deviation of the EW measurements for a line to scale the EWs of that data set, so that the two lines in comparison would appear to have the same amplitude of variability.  This is repeated with the line flux, and is done for both lines.  We then take the ratio of the normalized EWs for the two lines and the ratio of the normalized fluxes for the two lines.  We finally calculate the mean and standard deviation for these two sets of values.  As an example, consider Br $\gamma$ and H$_2$.  Let EW$_{NBr\gamma}$ = EW$_{Br\gamma}$/stdev(EW$_{Br\gamma}$), EW$_{NH2}$ = EW$_{N2}$/stdev(EW$_{H2}$), flux$_{NBr\gamma}$ = flux$_{Br\gamma}$/stdev(flux$_{Br\gamma}$), and flux$_{NH2}$ = flux$_{N2}$/stdev(flux$_{H2}$).  We calculate the mean and standard deviation of EW$_{NBr\gamma}$/EW$_{NH2}$ and flux$_{NBr\gamma}$/flux$_{NH2}$, where each of these ratios is a vector with one element per observation.

  If the variability of the lines is correlated and varies in a 1:1 way (i.e. when one line doubles in strength, the other doubles as well), then the EW and flux ratios will be 1 (due to the normalization by the standard deviation) and the standard deviation of the ratios will be 0.  It does not matter if one of the lines is intrinsically stronger than the other, just as long as the variability is a scaling of the two lines by the same factor.  If the variability of the lines are uncorrelated or anti-correlated, then the means and standard deviations will be random and are highly unlikely to be near 1 or 0, respectively.  This metric has several advantages.  The EW is unaffected by changes in extinction if the two lines are close, and the EW ratio is also similarly unaffected by changes in continuum veiling.  The line flux is unaffected by changes in continuum veiling, and the line flux ratio is also similarly unaffected by changes in extinction if the two lines are close in wavelength.  Whereas line flux is affected by non-photometric observing conditions and variable slit loss, the flux ratio of two lines (observed simultaneously) is not affected by these things, allowing us to make use of line flux data taken under non-photometric conditions.  Given uncorrelated input data, empirical testing showed that 95\% of simulated data sets had a standard deviation of ratios greater than 0.22.  Thus, if a real data set has a standard deviation less than 0.22, we considered it a real correlation using this metric, and found no correlation if the standard deviation was greater than 0.22.  This metric is insensitive to a negative correlation; as such a simulated data set with a negative correlation has a very large standard deviation.  

We used these two methods (comparing relative changes in EW, and the EW and flux ratios) to determine if there is a correlation between a pair of lines for a given target.  These two methods tend to give consistent results, giving conflicting results in 5/45 ($11\%$) of cases.  This suggests that comparing relative changes in EW is sufficient in most cases, and that the variability in the near-IR veiling is not sufficient to prevent using EW variability alone.  To test this, we plotted the relative change in the Ca I 2.264~$\mu$m feature and Br $\gamma$ (Figure 9).  The Ca I 2.264~$\mu$m feature is the only photospheric feature never to be seen in emission, and thus variability in the EW likely probes changes in K-band continuum veiling.  We find is no correlation between the changes in these two lines, showing that the measurement of the Br $\gamma$ emission line EW is not affected by changes in K-band continuum veiling.  This figure also shows that there is no close correlation between the mass accretion rate and near-IR continuum veiling on the time scales that we probed.  \citet{Gre2002} considered 3 possible sources of veiling emission: excess produced by disk, excess produced by the infalling envelope, and excess produced by infalling gas onto the star.  The K-band veiling is most sensitive to warm dust (in the disk or infalling envelope), but direct accretion of mass onto the star produces veiling at shorter wavelengths that may be better correlated with accretion diagnostics. 

Table 4 shows the results of applying these methods to our data set.  At the bottom of the table is an average score, intended to reflect the frequency of there being a positive correlation between a pair of lines.  A positive correlation is given 1 point, no correlation 0, and a negative correlation -1 points; these points then being averaged.

\subsection{Photometry}
    
   The K-band flux of our target stars and standard stars were measured using aperture photometry with the ATV program in IDL.  The aperture radius, inner and out sky radii were scaled with the FWHM (being 3, 6, and 9 times the FWHM, respectively) to ensure that the same fraction of the target's flux was in the aperture despite changes in the observing conditions.  The one exception was IRAS 04287+1081, which is spatially resolved.  In this case, we use an aperture radius, inner and outer sky radii of 20, 60, and 80 pixels respectively.

\section{Discussion}
   In this section, we discuss the results of our analysis of the variability of extinction, continuum veiling, photometry, and spectroscopic emission lines that probe mass accretion and winds.  We also compare the variability of mass accretion and wind tracers to determine if they are correlated.

\subsection{Extinction Variability}
   We estimated extinction using the [FeII] 1.256~$\mu$m  and 1.644~$\mu$m emission lines.  Since these lines share the same upper level and are optically thin, the observed line ratio can be compared to the intrinsic line ratio \citep{Con2010} to derive an extinction estimate.  There were 7 targets in our sample where both [FeII] lines were observed, and thus we were able to derive an extinction estimates using the [FeII] line ratio.   In all 7 cases ($100\%^{+0\%}_{-23\%}$) we observed statistically significant change in the extinction.  Since we saw a change in the extinction for all targets for which we estimated extinction, and since we expect that the circumstellar disks have a random orientation to our line of sight, we find that Class I protostars have significant non-homogeneous extinction at high latitudes above the circumstellar disk.  

   The extinction was seen to vary significantly on time scales ranging from days to years (see Figure 10).   We infer changes in A$_v$ extinction by up to 14 magnitudes within less than 10 days, and up to 17 magnitudes in 100 days. The shortest time interval over which we observed a change in the extinction was 2 days.  If we take 2 days as the orbital period of the extinguishing material, assuming a protostellar mass of 1~M$_{\odot}$ and radius of 1~R$_{\odot}$, then the extinguishing material is approximately 0.03~AU or about 6 stellar radii from the central protostar.  This suggests that, in most cases, the [FeII] shocked emission region is very close to the stellar surface.  IRAS 05256+3049 showed a particularly large change in extinction is a relatively short amount of time, e.g. the extinction changed by A$_v$=14.4 magnitudes within 10 days.  Our estimates show that a cloud that can add 14 magntiudes of visual extinction would be much larger than 0.03 AU, thus we believe that our line-of-sight is likely passing through a changing or rotating warped circumstellar disk.

    In contrast, the extinction of IRAS 20568+5217, which is an FU Ori-like star, changed relatively little, and then only on long time intervals.  The change in A$_v$ extinction is negligible on time intervals less than 100 days, but increases to 2.1 magnitudes on time intervals of 1 year, and 4.2 magnitudes on time intervals of 2 years.  This suggests that in this case the extinction to the [FeII] emission region is far from the central star.  Also, the extinction was quite low (the mean was A$_v=4.3$ mag, compared to a mean of A$_v=13.3$ mag for the sample as a whole), again suggesting that the [FeII] emission region is farther from the central star for FU Ori-like objects. 

   The median extinction change on time scales less than 30 days is about A$_v$=1 magnitude.  On timescales longer than 30 days, the change is about 2 magnitudes.  Figure 10 shows that the median extinction change on time scales greater than 1000 days is A$_v$=0.5 magnitudes, but this may be due to the smaller number of targets observed over such long time intervals. As it happened, the targets that showed strong extinction variability were not observed over the longest time scales.  We also note that most stars did not show a significant change in K-band brightness, which would seem to be inconsistent with the idea of all protostars having significantly variable extinction.  The one target (IRAS 03301+3111) for which we could measure both veiling and extinction showed no correlation between these values.  





\subsection{Veiling Variability}
    The near-IR continuum veiling (hereafter simply referred to as veiling) of young stars is caused by excess thermal emission from warm circumstellar material.  We were sensitive to changes in veiling for three targets (IRAS 03220+3035S, IRAS 03301+3111, and IRAS F23591+4748) that had apparent photospheric absorption lines (Figure 11).  We used the lines from Mg I at 1.488~$\mu$m, 1.503~$\mu$m (doublet), 1.711~$\mu$m, as well as the calcium doublet at 2.264~$\mu$m.  We did not use the 2.208 Na doublet or the CO band heads since the EWs of those features are affected by line emission from the disk.  Uncertainties in the photospheric spectral types of these stars lead to uncertainty in the veiling for any given star.  However, we were more sensitive to \textit{changes} in the veiling since changes in the EWs of the photospheric lines are easy to measure and is independent of the stellar spectral type.  To calculate the change in the veiling, we averaged the change in the EWs of these four photospheric lines.  The uncertainty in the veiling change was calculated using the standard deviation of change in the EWs of these four photospheric lines.  The mean variability of the veiling is 14.3\%, and the mean uncertainty in the change in veiling was 9.3\%.  This uncertainty is relatively small for IRAS 03220+3035S and IRAS 03301+3111 (5.0\% and 7.8\%, respectively), such that the amplitude of the variation is 2 to 3 times the uncertainty.  However, the uncertainty is much higher for IRAS F23591+4748 (14.9\%), such that the amplitude of the variation is less than the average uncertainty.  We observed statistically significant changes in the veiling for each of these three targets.  The shortest time interval found for a statistically significant change is 2 days (for IRAS 03301+3111). Similar to the above discussion of extinction variability, this implies that the warm material is only 0.03~AU or about 6 stellar radii away from the central protostar if on a Keplerian orbit.  This simple dynamical argument is within a factor of 2 of the expected dust sublimation radius for a 1~L$_{\odot}$ star \citep{Dul2007}.  We also note that the minimum time scale for changes in extinction and veiling are both limited by the minimum time span between our observations.   As mentioned above, Figure 9 shows that the veiling variability (as probed by changes in the EW of the Ca I 2.264~$\mu$m doublet) is uncorrelated with the Br $\gamma$ line EW, which may imply that the veiling variability is uncorrelated with the mass accretion rate \citep{Muz1998}.  



\subsection{Photometric Variability}
   The variability of the K-band photometry for our sample targets is shown in Figure 12.  For our sample targets, the standard deviations of the K-band photometry has a minimum of 0.03 magnitudes (IRAS 04287+1801, as we will see below, FU Ori-like stars tend to be remarkably stable), maximum of 0.45 magnitudes  (IRAS 18274$-$0212, the only target that showed a trend in its photometry), and a mean of 0.10 magnitudes.  In our sample, 15/19 targets have veiling high enough to prevent photospheric absorption lines to be seen in the K-band part of the spectrum.  Since the K-band flux for $\sim80\%$ of our targets is dominated by veiling flux, the photometric variability is likely driven by changes in veiling and/or changes in extinction (discussed below), rather than changes in the photospheric flux from the protostar.  A near-IR photometric survey of PMS stars in the $\rho$ Oph cloud \citep{Alv2008} found a mean peak to valley K-band variability of 0.15 magnitudes for member stars.  Our sample has a mean P-V K-band variability of 0.23 magnitudes; slightly greater variability than the older $\rho$ Oph sample of PMS stars.  
    
   Is the K-band photometric variability consistent with the observed variability in veiling and extinction?  The median variability of extinction (A$_v$ is 1.8 magnitudes, which is 0.19 magnitudes at K-band (Mathis 2000).  The mean change in the veiling, measured from 3 low-veiling targets, was 20\%, which is 0.20 magnitudes.  Assuming that the variability of the veiling and extinction are uncorrelated, we would expect the K-band photometric variability to be 0.28 magnitudes, roughly consistent with the observed variability.  The veiling variability was measured using only 3 targets that are not representative of the whole sample, being the only 3/19 targets with low veiling whereas the rest have high veiling.  Nevertheless, we find that the mean photometric variability of the high veiling targets ($0.10\pm0.10$) and low veiling targets ($0.08\pm0.03$) are consistent with each other.

   What is the characteristic time scale of photometric variability?  The rough, broken appearance of the K-band light curve suggest that the photometric variability is occurring on a time scale that is short compared to our observing cadence.  Observations from the YSOVAR project \citep{Mor2011} show that the average period for the mid-IR variability of their sample of YSOs is $\sim$6 days.  Since the time scale between our observations is typically much longer than this, the 6 day time scale is likely aliased in our K-band light curves.  




\subsection{Spectroscopic Variability Overview}

   Over the course of our monitoring program, the spectra of \emph{all} targets were observed to change.  Of the 19 targets, 15 ($79\%^{+10\%}_{-13\%}$) showed significant change within the shortest observing interval.  The spectra of all objects were found to vary significantly over periods of 39 days or less.  For the 8 targets where the minimum observing interval was 2 days or less, 7 targets ($88\%^{+10\%}_{-13\%}$) showed significant change within that time interval.  


   Despite the observation that the spectra of all targets were variable, there were no drastic changes in the spectra of any targets, with one exception.  IRAS 03301+3111 is the only target in the sample where the change in profile of the continuum of the spectrum is obvious upon visual inspection (Figures 13 and 14).  Otherwise, the spectra of the other target stars over all epochs are generally very similar, and the changes only become apparent when the spectra are compared in detail.  While many spectral features were observed to vary, few features were observed to completely vanish, or to become apparent when they were not previously apparent.  The spectra of targets with FU Ori-like spectra (deep water absorption bands, deep blue shifted He I absorption, few if any photospheric lines) have comparatively static spectra.  Although they have few emission or absorption lines in their spectra, the lines that they do have tend to be less variable than in the rest of the sample.


   In addition to the spectra of all targets being variable, all spectral lines are also variable, but not necessarily in all targets.  The EW of photospheric lines are variable due to changes in veiling (see Section 4.3).  We also noted that all emission lines, often associated with accretion or winds, are variable.  We measured the EW, line FWHM, and line center of the CO band head, as well as the He I, [FeII], H$_2$, Br $\gamma$ lines since monitoring accretion and wind tracers was a primary goal of this project.  Of the 8 targets with He I emission or absorption at 1.083~$\mu$m, we observed significant change in the EW in 7 targets ($88\%^{+11\%}_{-23\%}$).   Of the 9 targets with [FeII] emission at 1.644~$\mu$m, we observed significant change in the EW in 8 targets ($89\%^{+9\%}_{-21\%}$).   Of the 10 targets with H$_2$ emission at 2.121~$\mu$m, we observed significant change in the EW in all 10 targets ($100\%^{+0\%}_{-17\%}$).  \citet{Gre2010} also found variability in the H$_2$ S(1) 1-0 2.12~$\mu$m emission EW from night to night for Class I YSOs.

   For much of the following discussion, we use EW as a proxy to monitor changes in the line flux.  As shown above, veiling is variable, and can affect the EW measurement.  In section 3.3, we discuss how changes in veiling are unlikely to significantly affect EW measurements of emission lines.  Several figures show the variability of the EW of various lines versus time interval between observations.  Since the veiling has a mean variability of $37\%$, variability in these figures near or below this level may be due to the variable veiling and not due to changes in the lines themselves.  However, changes in the EW of these lines much greater than $37\%$ is most likely due to changes in the lines.  

   Many targets show a flat trend of variability versus time interval, suggesting that the variability on those time scales is sampling variability actually happening on a much shorter time scale.  We expect to see an increase in the median when there is an increase in the variability of a line at that time interval.  Whereas the median variability is often flat, for each line there is an 'envelope' of maximum variability that tends to increase with increasing time interval.  To understand the nature of the underlying light curves of these emission lines, we randomly sampled synthetic light curves to create plots similar to Figues 1-4.  Qualitatively, the best results were achieved by combining noise plus a sinusoidal term with roughly equal amplitude, and with the period of the sinusoidal term being about half of the base line of the observations.  SInce we sampled the actual emission line light curves on intervals from 2 days to 3 years, this result suggests that much of the variability happens on time scales much less than 2 days, and on a time scale of 1-3 years.

\subsection{Mass Accretion Tracers}
  H Br $\gamma$ is frequently used as a mass accretion tracer for PMS stars \citep{Muz1998}, but it has also been observed in the outflow \citep{Bec2010}.  15 targets showed HI Br $\gamma$ emission, and we observed statistically significant changes in the equivalent widths of all ($100\%^{+0\%}_{-12\%}$).  Figure 1 shows the relative variability of Br $\gamma$ as a function of the time interval between observations.  The median variability remains near $\sim15\%$ on time scales shorter than a hundred days, and increases to $\sim30\%$ on time scales of several hundred days.  This suggests that there are two primary time scales of mass accretion variability: variability on the time scale of days (or less), and variability on the time scale of a few years.  Variability on time scales of a day suggests a small spatial scale.  A Keplerian time of 1 day implies an orbital radius of 4 stellar radii (assuming solar values), corresponding to the inner edge of the accretion disk.    Variability on time scales of a few years corresponds to a spatial scale of 1 to 2 AU, suggesting that this variability traces the flow of matter onto the disk.  There is an apparent  'envelope' of variability, such that the maximum observed relative variability is proportional to the log of the time interval between observations.  Thus, the Br $\gamma$ line equivalent often varies by more than a factor of 3 on time scales longer than 1000 days.  An exception to this rule was a unique event observed from IRAS 23591+4748, where the Br $\gamma$ EW was observed to increase by a factor of $3.9\pm0.7$ within a span of two days.  

   We made a periodigram of the variability of the Br $\gamma$ emission (Figure 15), using the IDL program period.pro since our data are not evenly spaced in time.  Since each target has only a few epochs of observations, the periodigram for each target was very noisy.  Thus, we binned down the periodigrams and coadded the periodigrams for all 15 targets that showed Br $\gamma$ emission.  The resulting periodigram is nearly flat, suggesting that the Br $\gamma$ emission variability is similar to white noise.  There are no apparent spikes in frequency within the time scales sampled by these periodigrams.  With low confidence, there may be more variability on time scales of $\sim$30 days and $\sim$400 days than on time scales of $\sim$100 days.

   CO emission is also closely related to mass accretion in young stars, attributed to a hot upper layer within the inner 1 AU of the dense and hot accretion disk \citep{Cal1991}.  For this analysis, we only considered targets where CO was observed to be in emission, and measured the CO EW from 2.292~$\mu$m to 2.320~$\mu$m. \citet{Con2010} found that only Class I targets with high veiling have CO in emission, but not all high veiling targets have CO in emission.  Figure 2 shows the relative variability of CO as a function of the time interval between observations.  The median variability increases relatively gradually from $\sim5\%$ on time scales of a few days or less, to $\sim16\%$ on time scales of a 30 to 300 days, up to $\sim50\%$ on time scales of several years.  On time scales of less than 1 day, the CO variability (with one exception) never exceeds $30\%$.   The CO variability (with one exception) never exceeds $65\%$ on time scales of less than 1 year.  The lone exception is IRAS 21569+5842, which showed a change in the EW of a factor of 4 over a span of 30 days.  This corresponded with an increase in the line flux, suggesting that the change in EW was not simply due to a change in the continuum level.  On longer time scales, the relative variability exceeded a factor of 3.  As such, CO is most variable on a time scale of 1 to 3 years.  These time scales corresponds to a Keplerian disk radius of 1 AU to 2 AU.  

   The variability in the EW of these Br $\gamma$ and CO mass accretion tracers shows that mass accretion rates are variable.  We have estimated mass accretion rates only for those targets where we could estimate the extinction, which have Br $\gamma$ emission, and only for nights where we could flux calibrate the spectra.  To flux calibrate the spectra, we took K-band photometry of the target stars on our observing nights that were photometric.  The observed spectra were then scaled so that the integrated spectral flux density over the K-band filter band pass matched the observed flux of the target.  We corrected for extinction as described in Section 4.1.  The extinction was estimated for each target and for each observing session since extinction has been demonstrated to also be variable.  Our sample includes four targets (IRAS 03220+3035N, IRAS 03301+3111, IRAS 05256+3049, and IRAS 16442-0930) that satisfy all of these criteria, and these targets were observed on an average of 3 photometric nights each.  

   We used the results in \citet{Muz1998} to convert the Br $\gamma$ line flux to a mass accretion rate.  Fitting a linear regression line to their result, the best fit is $\log(\dot{M}(M_{\odot} \cdot yr^{-1})) = -8.85 + 0.55\log(F_{Br\gamma}(ergs \cdot s^{-1} \cdot cm^{-2}))$.  Since the targets in that paper are all in the Taurus star forming region, we were able to correct for the difference in distance between each of the four targets above and the Taurus star forming region by using the distances to the four targets listed in \citet{Con2007}.  The derived mass accretion rates versus time are shown in Figure 16.  The log of the mass accretion rates ranged from -8.5 $M_{\odot} \cdot yr^{-1}$ for IRAS 16442$-$0930 to -5.5 $M_{\odot} \cdot yr^{-1}$ for IRAS 03301+3111 and IRAS 05256+3049.  Standard deviation of mass accretion rate ranged from $\log(\dot{M}$) = 0.09 to 0.65 $M_{\odot} \cdot yr^{-1}$.  The average standard deviation is $\log(\dot{M}$) = 0.32 $M_{\odot} \cdot yr^{-1}$.

  The median of the Class I Br $\gamma$ derived mass accretion rates, shown in Figure 16, is $\sim$20 times higher than the median mass accretion rate for the T Tauri stars in \citet{Muz1998}, consistent with Br $\gamma$ emission tracing mass accretion among Class I YSOs.

   Here we consider two models for the excitation of CO emission.  \citet{Mar1997} proposed that the CO emission arises from the accretion column that is falling onto the star, similar to the Br $\gamma$ emission.  If this is correct, then we would expect a close correlation between the variability of the Br $\gamma$ and CO features, i.e. when one increases, the other should as well.   \citet{Naj1996} adopts a model, based on \citet{Cal1991}, with continuum from an optically thick disk mid-plane is seen through CO in an upper "line" layer.  If the mass accretion rate is extremely high, CO is seen in absorption (such as in FU Ori stars), whereas CO is seen in emission if the mass accretion rate is relatively low.  In this model, CO emission is from the inner 1~AU (whereas Br $\gamma$ originates in the accretion flow onto the star).  Since the emission region is on the order of 1~AU away from the central star, we would expect relatively little variability on time scales less than 1 year, and thus we would also expect a weak correlation (if any) between the variability of the Br $\gamma$ and CO features.  \citet{Con2010} showed that there is a correlation between the Br $\gamma$ and CO features in YSO spectra; specifically that CO is seen in emission only when Br $\gamma$ is also in emission.  Furthermore, CO is seen in absorption in excess of photospheric absorption only when Br $\gamma$ is not seen in emission.  

   In this data set, a few objects showed a weak correlation between Br $\gamma$ and CO variability, but more YSOs showed no correlation.  No object showed a negative correlation between Br $\gamma$ and CO variability.  As a group, no correlation is apparent (Figure 5).  The best correlation is for IRAS 05256+3049 and IRAS 21352+4307, where the data fall within two distinct clusters with no data points between these two clusters.  This suggests that the CO emission was either in a 'low' or 'high' state, and the Br $\gamma$ emission matched that state.  However, there is no correlation between smaller changes in Br $\gamma$ and CO variability on time scales of less than a year.  It seems that there is a correlation when there is a large change in CO emission, but not for smaller changes.  Furthermore, Figures 2 and 1 show that CO is not as variable on time scales of less than about a year, showing up to $\sim50\%$ variability whereas Br $\gamma$ shows over $250\%$ variability (each with a single but different exception).

    Table 4 shows that Br $\gamma$ and CO show positively correlated variability is less than half of the targets with these lines.  This further reinforces our finding that the excitation mechanism for these lines can be related, but otherwise do not seem to be closely coupled.  These observations are inconsistent with the model proposed by \citet{Mar1997}, and is more consistent with the model proposed by \citet{Cal1991}.  We observed larger CO variability on time scales longer than 1 year (Figure 2).  We also observed little correlation between the variability of Br $\gamma$ and CO on time scales much less than 1 year.  In summary, the data show that changes in the mass flow onto the stellar photosphere (traced by Br $\gamma$) and the disk surface temperature (traced by CO) are coupled for large changes on long time scales, but tend to vary independently on smaller scales and shorter time scales.

\subsection{Wind Tracers}
   [FeII] (1.644~$\mu$m), like all other emission lines, was observed to be variable.  The median variability of [FeII] was observed to increase with the time interval between observations, up to $35\%$.  The maximum observed variability 'envelope' also increased with time interval up to $\sim200\%$.  This is relatively low compared to the variability observed from Br $\gamma$ and CO.   The variability on time scales less than 30 days was also quite low, less than $25\%$.   

   The trends for the variability of H$_2$ (2.122~$\mu$m) are quite different from [FeII] (Figure 3).  The median variability versus time interval is relatively flat at $\sim20\%$.  This suggests that most of the variability of H$_2$ is on short time scales, and that the variability on long time scales primarily samples a faster variability.  The envelope of maximum variability versus time interval increases rapidly from $\>50\%$ on time scales of less than 10 days up to a factor of 11 change over $\sim800$ days.  However, only IRAS 21352+4307 shows H$_2$ variability greater than $100\%$.  Without this unusual object, the envelope would gradually increase to a maximum variability of $\sim100\%$, very similar to the envelope for [FeII].  
   
   [FeII] and H$_2$ emission lines are often observed together in the spectra of young stars.  \citet{Beck2008} concluded that H$_2$, similar to [FeII], is primarily shock excited among CTTS.  However, Among Class I YSOs, \citet{Gre2010} found that H$_2$ may be UV, X-ray, or shock excited.  It is possible for [FeII] and H$_2$ to be excited in different places in the wind, for several reasons (e.g. differences in excitation potential, temperature, density).  As an example, AO observations of BN/KL show a spatial separation between [FeII] and H$_2$ knots, with [FeII] emission mostly found near the tip of each 'bullet' and H$_2$ along the walls of the outflow.  Our previous efforts to use the ratio of the v=1-0 S(1) and v=2-1 S(1) lines, expected to be sensitive to the H$_2$ excitation mechanism, were inconclusive.

  If the [FeII] and H$_2$ emission lines share the same excitation mechanism for the targets in our sample, then we might expect that the [FeII] and H$_2$ lines should vary together.  Figure 6 shows the results of plotting the change in EW for [FeII] versus the change in the EW for H$_2$.  There is an overall positive correlation between changes in [FeII] and H$_2$.  Table 4 shows that the variability of  [FeII] and H$_2$ are frequently positively correlated.  For five targets, there is nearly a 1:1 relationship in the variability of [FeII] and H$_2$ (i.e. when the EW of one line doubles, so does the other), shown by the fact that the slope of the regression line for these five targets in Figure 6 is near 1.  This is consistent with what we would expect for a pair of lines that are both wind tracers and tend to both be excited by the same mechanism.  It would seem unlikely for a wind to excite these two lines in different places, under different physical conditions, and still show correlated variability.  We note that all of these objects have veiling that is high enough to mask photospheric absorption features.  For two targets (IRAS 03301+3111 and IRAS 16442$-$0930), the slope of the regression line is about 1/3, showing that H$_2$ changes by about a third of the change of [FeII].  This suggests that the H$_2$ emission is only partly excited by shocks, but mostly excited via another mechanism.  This cannot be explained by postulating that H$_2$ may be easier to excite for these targets, since for these stars, a 100\% increase in H$_2$ (for example) is matched with a 30\% increase in [FeII].  Both of these targets notably have low veiling.  The only target that had both lines and yet did not show a positive correlation is IRAS 21445+5712.  



   Our observation that the slope of the regression line in Figure 6 correlates with the veiling in the spectrum is a very important clue.  We speculate that high veiling targets have a hot inner disk, driving a strong or fast wind, resulting in the H$_2$ excitation being dominated by shocks.  In contrast, low veiling targets have a cool or evacuated inner disk, and drive a comparatively weak wind. They still have [FeII] and H$_2$ emission, but perhaps the wind is weaker or slower and thus the H$_2$ is mostly excited by something other than shocks.  This result is consistent with \citet{Gre2010} in that H$_2$ emission from Class Is YSOs have multiple excitation mechanisms.

   We also considered the variability in radial velocity the line center.  We would expect that the changes in the [FeII] and H$_2$ emission line radial velocities should follow each other if these lines originate in the same gas.  Changes in emission line radial velocity is independent of the line excitation mechanism(s).  In doing this analysis, we removed the radial velocity signature of the orbital motion of the Earth.  The uncertainty in our measurement of the line center is $\sim$3~kms$^{-1}$, not including systematic errors.  Figure 17 shows that there tends to be a positive correlation in the shifts in radial velocity of the [FeII] 1.64~$\mu$m and H$_2$ 2.12~$\mu$m emission lines the case for Class I YSOs.  In the figure, the dashed line has a slope of 1, which is the expected regression line if the radial velocities vary in step with each other.  We find that the median correlation coefficient is 0.55.  IRAS 04287+1801W and IRAS 21352+4307 have the highest correlation coefficients at 0.73 and 0.71, respectively.  For these two targets, it does seem that the radial velocity shifts are consistent with the [FeII] and H$_2$ emission originates in (kinematically) the same gas.  The lack of a strong correlation for the other targets...

   Observations and models (e.g. \citet{Edw2003}, \citet{Edw2006}, \citet{Kwa2007}) show that the He I (1.083~$\mu$m) absorption can originate in scattering by either a radially flowing stellar wind or a disk wind.  The He I is the most variable of any line with a median variability of $37\%$.  The line EW was observed to change by a factor of 12, and we also observed the line profile to vary as well.  Variability of up to $70\%$ was observed on a time scale as short as 2 days (Figure 4).  The Keplerian time of 2 days corresponds to and orbital radius of 0.03 AU, or 6 solar radii (assuming solar values), suggesting that the He I wind is launched very close to the stellar photosphere.  

   The He I emission is different for FU Ori-like YSOs and other non-FU Ori-like (i.e. 'regular') YSOs.  For the 'regular' YSOs, the He I line shows variable blue shifted and red shifted components, where either component can be in emission or absorption.  The EW of the He I line for these YSOs is highly variable, with a median variability of $57\%$, and a maximum variability of a factor of 12.   For objects with FU Ori-like spectra, the He I line is remarkably static.  The median variability is only $18\%$, and is flat versus time interval.  The maximum variability is less than $60\%$, and we only observed blue shifted absorption (with the exception of IRAS 04287+1801, which has blue shifted emission).  Since we see only a blue shifted component, then the far lobe is blocked from us, perhaps by the disk or envelope.  The blueshift of the line center ranges from -200~kms$^{-1}$ (IRAS 04287+1801W and IRAS 06297+1021W) to -350~kms$^{-1}$ (IRAS 20568+5217 and IRAS 21454+4718).  The velocity of the blue wing of the line spanned the range from -500~kms$^{-1}$ (IRAS 06297+1021W) to -780~kms$^{-1}$ (IRAS 20568+5217).  Other wind indicators, such as the [FeII] lines at 1.257~$\mu$m and 1.644~$\mu$m are also blue shifted.  The line centers of the [FeII] lines are blue shifted with a velocity of approximately -110~kms$^{-1}$, with the exception of IRAS 06297+1021W where the [FeII] line center is blue shifted by -275~kms$^{-1}$.  We note that while the [FeII] line is also blue shifted, it is a factor of 2-3 times slower than the wind traced by the He I line.  Compared to the maximum blue shifted wind velocity of the He I line from T Tauri stars (200-450 ~kms$^{-1}$), the winds from FU Ori-like stars (500-780~kms$^{-1}$) are much faster.

\subsection{Correlating Accretion and Wind Tracers}
   Perhaps the most important issue that these data can probe is the relationship between mass accretion and the wind.  The emission of [FeII], H$_2$, Br $\gamma$, and CO are all clearly related to the accretion process.  \citet{Con2010} showed that there is a connection between the maximum EWs of Br $\gamma$ and H$_2$ in Class I objects.  If the excitation mechanisms of these lines are directly related and the lines form at similar times, then we would expect a positive correlation in their variability.  

   Plots of the relative change in [FeII] or H$_2$ versus change in Br $\gamma$ show a wide variety of relationships (Figures 7 and 8).  For most objects, there is no clear trend.  For the objects that do show a correlation between the changes in accretion and wind tracers, some (IRAS 05256+3049, 21352+4307, 03301+3111) show a positive correlation, whereas an equal number of YSOs (IRAS 04239+2436, 21569+5842, 16442$-$0930) show a negative correlation.  Whether the YSO has a positive or negative correlation between accretion and wind tracers does not depend on the veiling or whether CO is in emission or absorption (which itself is correlated with the veiling).  We also addressed the question of whether the likelihood of a correlation is related to the emission line strength.  We compared the Student's t statistic (Table 3) for Br $\gamma$ vs CO and for Br $\gamma$ vs [FeII] with the EW and flux for those lines.  We found no correlation between the t statistic and the EW nor the fluxes.  

   Figure 18 shows the correlation coefficients when the relative change of various accretion and wind tracers are plotted against each other.   We stress that the EWs of these lines were extracted from the same spectra, and these lines were observed simultaneously.  When comparing the EW change of Br $\gamma$ versus CO emission, there is occasionally a positive correlation, but often there is no correlation.  It is important to note that we have not found a negative correlation between these two lines.  H$_2$ versus [FeII] often shows a strong positive correlation, and again we note that we have not found a negative correlation between these two lines.  We found positive and negative correlations when comparing accretion tracers (Br $\gamma$ or CO) versus wind tracers (H$_2$ or [FeII]), showing that there is a complex relationship between accretion and the wind.  



 Table 3 lists whether there is an actual correlation between these lines with a confidence greater than 95$\%$ based on the relative variability of EW.  The correlation confidence was computed by using the linear correlation coefficient in the t test \citep{Wal2008}. Table 4 shows the correlations for each target across rows, and the correlation for each pair of lines down the columns.  Table 4 reinforces our finding that comparing the relative change of Br $\gamma$ vs. CO and H$_2$ vs. [FeII] only finds positive correlation or no correlation, but never finds a negative correlation.  Also, we found positive and negative correlations when comparing accretion tracers vs. wind tracers.  This can also be consistent with accretion powering the wind if the line emissions triggered by winds are separated from the accretion tracers by a large amount of time (larger than between Br $\gamma$ and CO emissions).



   We have established that there appears to be a complex relationship between accretion and wind tracers, with some targets showing a positive correlation whereas other targets show a negative correlation.  To further explore this, we plotted the relative change of Br $\gamma$ vs. H$_2$ vs. [FeII] in 3-space.  An example of such a plot is shown in Figure 19.  There are five targets in our sample that show Br $\gamma$ and. H$_2$ and [FeII] in emission.  We surprisingly found that the relative change of these lines \textit{always} fell along a tightly defined plane.  The orientation of this plane varied greatly from target to target.  One hypothesis is that the data lie on a plane simply because H$_2$ and [FeII] tend to be well correlated but Br $\gamma$ is uncorrelated with either, in which case we would expect that the plane made by the data would be perpendicular to the H$_2$ vs. [FeII] plane.  However, this is not the case; the best results are found by rotating the 3-space such that the H$_2$ vs. [FeII] plane is not in the plane of the page.  Furthermore, the orientation of the best fit plane varies from target to target.  It may be possible for veiling variability to produce a planar relationship between the EW changes of these three lines.  However, if that were the case, then we would expect much stronger correlations between the variability of Br $\gamma$ and H$_2$, which are very close in wavelength.

     We fit a plane to the data, using the equation $\Delta$Br$\gamma$ = R$_0$ + R$_1$$\Delta$[FeII] +  R$_0$$\Delta$H$_2$, where $\Delta$ is the relative change in the line EW.  The coefficients of the fit are listed in Table 5.  This result again suggests that there is a strong connection between accretion and the wind, but that connection is complex and is different for individual YSOs.  One might suspect that such a planar relationship is an artifact of a systematic error.  As discussed above, variability in the veiling does not tend to affect the ability of EW to characterize correlations between lines.  Also, if veiling variability were the cause of this planar relationship, then the orientations of the planes in 3-space would be the same.  We encourage theorists to explore whether these observations are consistent with plausible physical origins of these line emissions in these individual objects. 


   Whereas the variability of wind tracers ([FeII] and H$_2$) are often correlated, and the variability of the accretion tracers Br $\gamma$ are occasionally correlated, Table 4 shows that the variability of the wind and accretion tracers are rarely correlated.  Only 2/8 targets show a positive correlation between Br $\gamma$ and H$_2$, and 2/7 targets show a positive correlation between CO and H$_2$.  Only 1/6 targets show a positive correlations with Br $\gamma$ or CO with [FeII].  In both cases, IRAS 03300+3111 has the positive correlation, suggesting that this positive correlation is rare and that this may be an anomalous target.  The difference in the number of targets with positive correlations between H$_2$ and accretion tracers versus [FeII] and accretion tracers is too small to suggest that there is truly a difference between how these two wind tracers correlate with accretion.  Whereas the wind is clearly driven by the accretion process, the data show that the variability of the accretion tracers and wind tracers tend not to be closely correlated on time scales from days to years.  

\section{Comments on Individual Objects}
   We now note alternative designations or particularly interesting properties of the individual objects we observed. Most objects do have properties or identifications worth noting, but we have no noteworthy comments for IRAS 05256+3049, IRAS 16442−0930, or IRAS 18275+0040.

\textbf{IRAS 03220+3035 (N)}  L1448 IRS 1.  1\farcs4 binary with IRAS 03220+3035 (S).  

\textbf{IRAS 03220+3035 (S)}  We observed a significant change in the veiling for this target.  The average veiling is r$_k=0.44\pm0.44$.  The veiling changed by 37\%, with an average uncertainty in the relative veiling of 6.5\%.  There were three intervals over which the change in veiling was statistically significant.

\textbf{IRAS 03301+3111}   In Barnard 1.  This is the only target with positive correlations between all accretion/wind tracer pairs that were analyzed.  This is the only target in the sample where the continuum of the spectrum significantly changed over the course of this program. In this case, the veiling increased from 0.01$\pm$0.10 in Nov 2006 to 0.45$\pm$0.07 in Aug 2009.  This change was accompanied by the photospheric CO absorption being pushed in emission, and the strengthening of the emission lines from H, H$_2$, and Ca.  The change in veiling was also accompanied by a change in the overall slope of the spectrum at K-band (see Figure 13).  Figure 14 more clearly shows the change in the spectrum, and especially the water emission from $\sim$1.7~$\mu$m to $\sim$2.1~$\mu$m.  This is also the only target to /emph{not} show a clear correlation between the radial velocity shifts of the [FeII] and H$_2$ emission lines.  IRAS 03301+3111 appears to be an exceptional object, whose accretion and outflow properties seem to be different than for other Class I YSOs.  

   We propose the following scenario to explain these changes.  In 2006, this protostar had a relatively low accretion rate and little dust near the star, similar to a T Tauri star but with higher exinction.  By 2009, a significant amount of circumstellar material (including dust and water) had moved closer to the star and was heated, accounting for the increased veiling and water emission.  Accretion also increased, leading to increased irradiation of the disk surface.  As the temperature of the disk surface had exceeded the disk midplane, we saw CO in emission.  In \citet{Con2010}, 20/110 (18\%$\pm$4\%) objects have a similar spectrum (late type continuum, CO absorption, H and often H$_2$ emission).  This suggests that, while such dramatic changes in the spectrum are rare, objects that might undergo such a change may be quite common.  

\textbf{IRAS 04239+2436} HH 300 VLA 1

\textbf{IRAS 04287+1801 (W)} L1551 IRS 5B, a well known YSO with an FU Ori-like spectrum.


\textbf{IRAS 05513$-$1024}  An early type star with strong veiling.

\textbf{IRAS 06297+1021 (W)}  Similar to an FU Ori star, the spectrum shows strong water vapor absorption and strong HeI absorption.  However, CO and many other lines are seen in emission.  

\textbf{IRAS 16240$-$2430 (W)} Likely an early type star with strong veiling.  Located in L1681


\textbf{IRAS 18274$-$0212}  The only target to show a trend in its K-band photometry.  Spectrum shows strong veiling and strong H$_2$ emission.


\textbf{IRAS 20377+5658} Located in L1036.

\textbf{IRAS 20568+5217} HH 381 IRS.  Star with FU Ori-like spectrum.

\textbf{IRAS 21352+4307}   This object also showed uniquely high H$_2$ variability (Figure 3), and is the only object to show H$_2$ variability greater than 100\%, with the H$_2$ EW increasing by a factor of 10 within 2 years.  

\textbf{IRAS 21445+5712}  IC 1396 East.  This is the only target with [FeII] and H$_2$ where the variability of these lines was uncorrelated.  In this case, the [FeII] varied but the H$_2$ was remarkably static.  

\textbf{IRAS 21454+4718}  V1735 Cyg.  FU-Ori like spectrum

\textbf{IRAS 21569+5842} In L1143.  The CO emission EW changed by a factor of 4 in 32 days.

\textbf{IRAS F23591+4748}  We observed the primary component of this binary protostar.  We observed a significant change in the veiling for this target.  The average veiling is r$_k=0.76\pm0.34$.  The veiling changed by 29\%, with an average uncertainty in the relative veiling of 15\%, over a time period of $<100$ days.  The Br $\gamma$ line flux and EW, tracing mass accretion rate, increased by a factor of 4 in two days.  

\section{Summary and Conclusions}
  We observed the near-IR spectra of a sample of 19 embedded (Class I) protostars over time scales ranging from 2 days to 3 years, spanning 4 -9 epochs each. We conclude the following:

1) Every target had variable spectroscopic features, and each spectroscopic feature was observed to vary, but not in every target.

2) Although the spectrum of each target was variable, there were no drastic changes in the continuum (with one exception), nor did any lines disappear or appear when the line was not previously visible.

3) The median amplitude variability of Br $\gamma$ versus time interval is flat on time scales less than $\sim$100 days, suggesting that the variability seen on time scales less than 100 days is sampling a variability that is happening on a time scale similar to, or shorter than, the 2 day minimum interval between observations.  

4) Extinction was measured from the ratio of 2 [FeII] lines and was found to be variable on time scales less than 10 days.  This demonstrates that the shocks that excite the [FeII] emission that we observe are very close to the stellar surface, and that there are also high opacity clouds similarly close to the star.  We also found continuum veiling to be variable on similarly short time scales.

5) The variability of [FeII] and H$_2$ wind diagnostics are closely correlated for most targets, suggesting that these two lines usually have the same excitation mechanism.  This is not true for all targets, showing that the excitation mechanisms for these two lines can be different.   Br $\gamma$ and CO line emissions are also positively correlated in some objects but not in others. Although both lines probe accretion, this result may be consistent with these lines forming at different distances from the central protostar, and being excited in different ways.

6) The variability of wind and accretion diagnostics are rarely positively correlated, suggesting a significant decoupling in time between the accretion process and the modulation of the wind emissions.  

7) Compared to photometric studies of T Tauri stars, our Class I YSOs show a higher amplitude of variability, and roughly the amount of variability expected considering the observed variability of veiling and extinction.  Class I YSOs have higher extinction and veiling overall compared to T Tauri stars, so the increased photometric variability is expected.

\acknowledgments
We are grateful for the professional assistance from Bill Golish, Dave Griep, Paul Sears, and Eric Volquardsen at the IRTF.  This research has made use of the SIMBAD database, operated at CDS, Strasbourg, France, and NASA's Astrophysics Data System.  This publication makes use of data products from the Two Micron All Sky Survey, which is a joint project of the University of Massachusetts and the Infrared Processing and Analysis Center/California Institute of Technology, funded by the National Aeronautics and Space Administration and the National Science Foundation.  This research has made use of NASA's Astrophysics Data System.  This research was supported by an appointment to the NASA Postdoctoral Program at the Ames Research Center, administered by the Oak Ridge Associated Universities through a contract with NASA.



Facilities: \facility{IRTF}.

\clearpage


\clearpage
\begin{deluxetable}{lccccccccccccc}
\tabletypesize{\scriptsize}
\tablecaption{Source Characteristics}
\tablewidth{0pt}
\tablecolumns{6}
\tablehead{
\colhead{IRAS} &
\colhead{$\alpha$(J2000)\tablenotemark{a}} &
\colhead{$\delta$(J2000)\tablenotemark{a}} & 
\colhead{K$_s$\tablenotemark{b}} & 
\colhead{$\alpha$\tablenotemark{c}} &
\colhead{L$_{bol}$\tablenotemark{d} (L$_{\odot}$)} &
}
\startdata   
03220+3035(N)     & 03 25 09.43       &   +30 46 21.6   & 10.83       &  0.02   &     2.0    \\ 
03220+3035(S)     & 03 25 09.43       &   +30 46 21.6   & 10.57       &  0.02   &  \nodata   \\ 
03301+3111        & 03 33 12.84       &   +31 21 24.1   & 10.58       &  0.31   &     4.0    \\ 
04239+2436        & 04 26 56.30       &   +24 43 35.3   & 10.22       &  0.09   &     1.1    \\ 
04287+1801        & 04 31 34.08       &   +18 08 04.9   &  9.25$^{1}$ &  0.76   &    20.2    \\ 
05256+3049        & 05 28 49.86       &   +30 51 29.3   & 10.17       &  0.16   &  6417.3    \\ 
05513-1024        & 05 53 42.55       & $-$10 24 00.7   &  5.96$^{1}$ &  0.18   &   \nodata  \\ 
06297+1021(W)     & 06 32 26.12       &   +10 19 18.4   &  8.14       & \nodata &  \nodata   \\ 
16240$-$2430(W)   & 16 27 02.34       & $-$24 37 27.2   &  7.82       & \nodata &  \nodata   \\ 
16442$-$0930      & 16 46 58.27       & $-$09 35 19.7   & 10.92$^{1}$ &  0.22   &      0.7   \\ 
18274$-$0212      & 18 30 01.36       & $-$02 10 25.6   & 11.66       &  0.12   &  \nodata   \\ 
18275+0040        & 18 30 06.17       &   +00 42 33.6   &  7.72       & -0.19   &     3.4    \\ 
20377+5658        & 20 38 57.48       &   +57 09 37.6   &  9.48       &  0.32   &      4.8   \\ 
20568+5217        & 20 58 21.09       &   +52 29 27.7   &  7.90       &  0.62   &     45.6   \\ 
21352+4307        & 21 37 11.39       &   +43 20 38.4   & 11.75       &  0.17   &     11.7   \\ 
21445+5712        & 21 46 07.12       &   +57 26 31.8   & 10.25       &  0.54   &     18.5   \\ 
21454+4718        & 21 47 20.66       &   +47 32 03.6   &  7.03       &  0.70   &    106.7   \\ 
21569+5842        & 21 58 35.90       &   +58 57 22.8   & 10.25       &  0.08   &      1.0   \\ 
F23591+4748       & 00 01 43.25       &   +48 05 19.0   & 10.42       &  0.60   &   \nodata  \\ 
\enddata

\tablenotetext{a}{ 2MASS coordinate for candidate YSO.}
\tablenotetext{b}{ K-band magnitudes from Table 4 from Connelley et al. (2008), and are in the MKO photometric system unless otherwise noted.  Magnitudes noted with $^{1}$ are from 2MASS and are in the 2MASS photometric system.}
\tablenotetext{c}{ $\alpha$ is the spectral index of the source (Lada 1991), and is calculated from the IRAS fluxes of the source.}
\tablenotetext{d}{ Bolometric luminosity, from \citet{Con2008} }
\end{deluxetable}

\clearpage
\begin{deluxetable}{lcccccccccccccccccccccc}
\rotate
\setlength{\tabcolsep}{0.04in}
\tabletypesize{\scriptsize}
\tablecaption{Observing Log}
\tablewidth{0pt}
\tablecolumns{20}
\tablehead{
\multicolumn{1}{c}{UT Date} &
\multicolumn{1}{c}{03220} &
\multicolumn{1}{c}{03220} & 
\multicolumn{1}{c}{03301} & 
\multicolumn{1}{c}{04239} &
\multicolumn{1}{c}{04287} &
\multicolumn{1}{c}{05256} &
\multicolumn{1}{c}{05513} &
\multicolumn{1}{c}{06297} &
\multicolumn{1}{c}{16240} &
\multicolumn{1}{c}{16442} &
\multicolumn{1}{c}{18274} &
\multicolumn{1}{c}{18275} &
\multicolumn{1}{c}{20377} &
\multicolumn{1}{c}{20568} &
\multicolumn{1}{c}{21352} &
\multicolumn{1}{c}{21445} &
\multicolumn{1}{c}{21454} &
\multicolumn{1}{c}{21569} &
\multicolumn{1}{c}{F23591} \\
\multicolumn{1}{c}{ } &   
\multicolumn{1}{c}{+3035N} &
\multicolumn{1}{c}{+3035S} & 
\multicolumn{1}{c}{+3111} & 
\multicolumn{1}{c}{+2436} &
\multicolumn{1}{c}{+1801W} &
\multicolumn{1}{c}{+3049} &
\multicolumn{1}{c}{$-$1024} &
\multicolumn{1}{c}{+1021W} &
\multicolumn{1}{c}{$-$2430W} &
\multicolumn{1}{c}{$-$0930} &
\multicolumn{1}{c}{$-$0212} &
\multicolumn{1}{c}{+0040} &
\multicolumn{1}{c}{+5658} &
\multicolumn{1}{c}{+5217} &
\multicolumn{1}{c}{+4307} &
\multicolumn{1}{c}{+5712} &
\multicolumn{1}{c}{+4718} &
\multicolumn{1}{c}{+5842} &
\multicolumn{1}{c}{+4748} \\
}
\startdata   
2006 Oct 31  &        &        &        &  s     &        &        &        &        &        &        &        &        &        &        &        &        &        &        &        \\
2006 Nov 01  &        &        &        &        &        &        &  s     &        &        &        &        &        &        &        &        &        &        &        &        \\
2006 Nov 15  &        &        &        &        &  s,p   &  s,p   &        &        &        &        &        &        &        &        &        &        &        &        &        \\
2006 Nov 16  &  s,p   &  s,p   &        &        &        &        &        &  s,p   &        &        &        &        &        &        &        &        &        &        &        \\
2006 Nov 17  &        &        &  s,p   &        &        &        &        &        &        &        &        &        &        &        &        &        &        &        &        \\
2007 Jul 12  &        &        &        &        &        &        &        &        &        &        &        &        &        &  s     &        &        &        &        &  s     \\
2007 Jul 13  &        &        &        &        &        &        &        &        &  s     &        &  s     &        &  s     &        &        &  s     &  s     &  s     &        \\
2007 Jul 14  &        &        &        &        &        &        &        &        &        &  s     &        &  s     &        &        &  s     &        &        &        &  s     \\
2007 Oct 12  &        &        &        &        &        &        &        &        &        &        &  s,p   &        &        &        &        &  s,p   &        &        &  s,p   \\
2007 Dec 18  &        &        &        &        &        &  s     &        &  s     &        &        &        &        &        &        &        &  s     &        &        &        \\
2008 Jun 26  &        &        &        &        &        &        &        &        &        &  s,p   &        &  s     &  s     &  s     &  s     &        &        &  s     &        \\
2008 Jul 28  &        &        &        &        &        &        &        &        &        &        &        &        &  s     &        &  s     &        &        &  s     &  s     \\
2009 Apr 13  &        &        &        &        &        &        &        &        &  s     &        &        &  s     &        &        &        &        &        &        &        \\
2009 Apr 14  &        &        &        &        &        &        &        &        &  s     &  s     &        &        &        &        &        &        &        &        &        \\
2009 May  1  &        &        &        &        &        &        &        &        &  s,p   &  s,p   &  s,p   &  s,p   &  s,p   &  s,p   &  s,p   &  s,p   &  s,p   &        &        \\
2009 Jun 11  &        &        &        &        &        &        &        &        &  s,p   &  s,p   &  s,p   &  p     &  s,p   &        &  s,p   &        &        &        &        \\
2009 Jun 12  &        &        &        &        &        &        &        &        &        &        &        &  s,p   &        &  s,p   &        &  s,p   &  s,p   &  s,p   &  s,p   \\
2009 Jun 25  &        &        &        &        &        &        &        &        &        &        &  s,p   &  s,p   &  s,p   &        &        &  s,p   &  s,p   &        &        \\
2009 Jul 19  &        &        &        &        &        &        &        &        &  s,p   &        &  s,p   &        &        &  s,p   &        &        &  s,p   &        &        \\
2009 Jul 20  &        &        &        &        &        &        &        &        &  s     &  s     &  s     &  s     &  s     &        &  s     &  s     &        &        &        \\
2009 Aug 19  &  s,p   &  s,p   &  s,p   &  s,p   &  s,p   &        &        &        &        &        &        &        &        &        &        &        &        &  s,p   &  s,p   \\
2009 Aug 31  &  s,p   &  p     &  s,p   &  s,p   &  s,p   &  s,p   &        &        &        &        &        &        &        &        &        &        &        &        &  s,p   \\
2009 Sep 05  &  s,p   &  s,p   &  s,p   &  s,p   &  s,p   &  s,p   &        &  s,p   &        &        &        &        &        &        &        &        &        &  s,p   &  s,p   \\
2009 Sep 07  &  s     &        &  s     &  s     &        &  s     &  s     &        &        &        &        &        &        &        &        &        &        &        &        \\
2009 Sep 21  &  s,p   &  p     &  s,p   &  s,p   &  s,p   &  s,p   &  s,p   &  s,p   &        &        &        &        &        &        &        &        &        &  s,p   &  s,p   \\
2010 Jan 14  &  s     &  s     &  s     &  s     &  s     &        &        &  s     &        &        &        &        &        &        &        &        &        &        &        \\
2010 Dec 07  &        &        &        &        &        &        &        &  s     &        &        &        &        &        &        &        &        &        &        &        \\
2010 Dec 09  &  s,p   &  s,p   &  s     &  s,p   &  s     &  s     &  s     &  s     &        &        &        &        &        &        &        &        &        &        &        \\

\enddata

\tablecomments{s: Spectra, p: Photometry }

\end{deluxetable}

\clearpage
\begin{deluxetable}{llcccccccccccc}
\tabletypesize{\scriptsize}
\tablecaption{Emission Line Variability Correlations}
\tablewidth{0pt}
\tablecolumns{7}
\tablehead{
\colhead{IRAS} &
\colhead{Line Ratio} &
\colhead{N\tablenotemark{a}} & 
\colhead{r\tablenotemark{b}} & 
\colhead{t\tablenotemark{c}} &
\colhead{Correlation\tablenotemark{d}} &
\colhead{Linear Fit Slope} &
}
\startdata   
03220+3035N   &    Br$\gamma$/CO  &   28   &    +0.22   &   +1.15  &  none  &  $+0.14\pm0.12$ \\
              & Br$\gamma$/[FeII] &   28   &  $-$0.25   & $-$1.31  &  none  &  $-0.21\pm0.16$ \\
              &         CO/[FeII] &   28   &  $-$0.26   & $-$1.37  &  none  &  $-0.34\pm0.25$ \\
03301+3111    &    Br$\gamma$/CO  &   21   &    +0.43   &   +2.08  &  +     &  $+0.68\pm0.32$ \\
              & Br$\gamma$/H$_2$  &   28   &    +0.59   &   +3.73  &  +     &  $+0.94\pm0.44$ \\
              & Br$\gamma$/[FeII] &   28   &    +0.97   &  +19.05  &  +     &  $+1.35\pm0.14$ \\
              &         CO/[FeII] &   21   &    +0.45   &   +2.25  &  +     &  $+0.43\pm0.19$ \\
              &      H$_2$/[FeII] &   28   &    +0.66   &   +4.48  &  +     &  $+0.40\pm0.13$ \\
              &          CO/H$_2$ &   21   &    +0.44   &   +2.17  &  +     &  $+0.48\pm0.23$ \\
04239+2436    &    Br$\gamma$/CO  &   28   &  $-$0.20   & $-$1.04  &  none  &  $-0.08\pm0.07$ \\
              & Br$\gamma$/H$_2$  &   28   &  $-$0.82   & $-$7.31  &  $-$   &  $-0.35\pm0.05$ \\
              & Br$\gamma$/[FeII] &   28   &  $-$0.24   & $-$1.26  &  none  &  $-0.12\pm0.09$ \\
              &         CO/[FeII] &   28   &    +0.17   &   +0.89  &  none  &  $+0.23\pm0.25$ \\
              &      H$_2$/[FeII] &   28   &    +0.67   &   +4.60  &  +     &  $+0.77\pm0.18$ \\
              &          CO/H$_2$ &   28   &    +0.27   &   +1.43  &  none  &  $+0.30\pm0.21$ \\
04287+1801W   &      H$_2$/[FeII] &   21   &    +0.90   &   +9.00  &  +     &  $+0.97\pm0.10$ \\
05256+3049    &    Br$\gamma$/CO  &   21   &    +0.36   &   +1.68  &  none  &  $+0.29\pm0.18$ \\
              & Br$\gamma$/H$_2$  &   21   &    +0.55   &   +2.87  &  +     &  $+0.70\pm0.25$ \\
              & Br$\gamma$/[FeII] &   21   &    +0.23   &   +1.03  &  none  &  $+0.17\pm0.25$ \\
              &         CO/[FeII] &   21   &  $-$0.63   & $-$3.54  &  $-$   &  $-0.82\pm0.23$ \\
              &      H$_2$/[FeII] &   21   &    +0.84   &   +6.75  &  +     &  $+0.66\pm0.12$ \\
              &          CO/H$_2$ &   21   &  $-$0.54   & $-$2.80  &  none  &  $-0.82\pm0.30$ \\
05513$-$1024    &    Br$\gamma$/CO  &    6   &    +0.39   &   +0.88  &  none  &  $+0.33\pm0.34$ \\
06297+1021W   &    Br$\gamma$/CO  &   21   &    +0.69   &   +4.16  &  +     &  $+1.58\pm0.32$ \\
16240$-$2430W   &    Br$\gamma$/CO  &   21   &    +0.64   &   +3.63  &  +     &  $+0.49\pm0.14$ \\
16442$-$0930    & Br$\gamma$/H$_2$  &   15   &  $-$0.52   & $-$2.19  &  $-$   &  $-1.50\pm0.67$ \\
              & Br$\gamma$/[FeII] &   15   &  $-$0.81   & $-$5.02  &  $-$   &  $-0.99\pm0.26$ \\
              &      H$_2$/[FeII] &   15   &    +0.63   &   +2.92  &  +     &  $+0.30\pm0.11$ \\
18274$-$0212    &    Br$\gamma$/CO  &   21   &    +0.30   &   +1.37  &  none  &  $+0.30\pm0.24$ \\
              & Br$\gamma$/H$_2$  &   21   &    +0.52   &   +2.65  &  +     &  $+0.58\pm0.23$ \\
              &          CO/H$_2$ &   21   &    +0.32   &   +1.46  &  none  &  $+0.34\pm0.23$ \\
20377+5658    &    Br$\gamma$/CO  &   21   &    +0.52   &   +2.72  &  +     &  $+1.04\pm0.39$ \\
21352+4307    &    Br$\gamma$/CO  &   10   &    +0.61   &   +2.18  &  +     &  $+0.77\pm0.35$ \\
              & Br$\gamma$/H$_2$  &   10   &  $-$0.05   & $-$0.15  &  none  &  $-0.10\pm0.67$ \\
              & Br$\gamma$/[FeII] &   10   &  $-$0.50   & $-$1.16  &  none  &  $-0.73\pm0.45$ \\
              &         CO/[FeII] &   10   &  $-$0.74   & $-$3.11  &  $-$   &  $-0.16\pm0.05$ \\
              &      H$_2$/[FeII] &   10   &    +0.88   &   +5.24  &  +     &  $+1.11\pm0.21$ \\
              &          CO/H$_2$ &   10   &  $-$0.69   & $-$2.70  &  $-$   &  $-0.12\pm0.04$ \\
21445+5712    &      H$_2$/[FeII] &   15   &  $-$0.04   & $-$0.15  &  none  &  $-0.01\pm0.04$ \\
21569+5842    &    Br$\gamma$/CO  &   21   &  $-$0.13   & $-$0.59  &  none  &  $-0.38\pm0.64$ \\
              & Br$\gamma$/H$_2$  &   21   &  $-$0.57   & $-$3.02  &  $-$   &  $-0.61\pm0.20$ \\
              &          CO/H$_2$ &   21   &    +0.49   &   +2.42  &  +     &  $+1.78\pm0.74$ \\
F23591+4748   & Br$\gamma$/H$_2$  &   21   &  $-$0.31   & $-$1.42  &  none  &  $-0.24\pm0.21$ \\

\enddata
\tablenotetext{a}{Number of data points}
\tablenotetext{b}{Correlation Coefficient}
\tablenotetext{c}{Value for Student's t test.  If N=21, t$>$2.08 for two-tailed significance to be 0.050.  If N=28, t$>$2.04}
\tablenotetext{d}{Confidence criteria is $95\%$.  none: No correlation; +: positive correlation; $-$: negative correlation}

\end{deluxetable}

\clearpage
\begin{deluxetable}{lccccccccccccc}
\tabletypesize{\scriptsize}
\tablecaption{Emission Line Variability Correlations}
\tablewidth{0pt}
\tablecolumns{7}
\tablehead{
\colhead{IRAS} &
\colhead{Br$\gamma$/CO} &
\colhead{Br$\gamma$/H$_2$} & 
\colhead{Br$\gamma$/[FeII]} & 
\colhead{CO/[FeII]} &
\colhead{H$_2$/[FeII]} &
\colhead{CO/H$_2$} &
}
\startdata   
03220+3035(N)     &  N,N     &            &  N,N      &  N,N     &           &          \\ 
03220+3035(S)     &          &            &           &          &           &          \\ 
03301+3111        &  +,+     &  +,+       &  +,+      &  +,+     &  +,+      &  +,+     \\ 
04239+2436        &  N,N     &  $-$,N     &  N,N      &  N,+     &  +,+      &  N,+     \\ 
04287+1801        &          &            &           &  N,N     &  +,+      &  N,N     \\ 
05256+3049        &  N,N     &  +,N       &  N,N      &  $-$,N   &  +,+      &  N,N    \\ 
05513$-$1024      &  N,N     &            &           &          &           &          \\ 
06297+1021(W)     &  +,+     &            &           &          &           &          \\ 
16240$-$2430(W)   &  +,+     &            &           &          &           &          \\ 
16442$-$0930      &          &  $-$,+     &  $-$,N    &          &  +,+      &          \\ 
18274$-$0212      &  N,N     &  +,+       &           &          &           &  +,N     \\ 
18275+0040        &          &            &           &          &           &          \\ 
20377+5658        &  +,+     &            &           &          &           &          \\ 
20568+5217        &          &            &           &          &           &          \\ 
21352+4307        &  +,N     &  N,N       &  N,N      &  $-$,N   &  +,+      &  $-$,N   \\ 
21445+5712        &          &            &           &          &  N,N      &          \\ 
21454+4718        &          &            &           &          &           &          \\ 
21569+5842        &  N,N     &  $-$,N     &           &          &           &  +,+     \\ 
F23591+4748       &          &  N,N       &           &          &           &          \\ 
Average           &  $0.41\pm0.14$  &   $0.19\pm0.11$     &   $0.08\pm0.08$    &  $0.08\pm0.08$   &   $0.86\pm0.25$    &   $0.36\pm0.16$   \\  
\enddata

\tablecomments{N: no correlation, +: positive correlation with greater than 95\% confidence, -: negative correlation with greater than 95\% confidence.  The result is in the form of x,y where x is the result from plotting the relative variability of the EWs whereas y is the result from the metric discussed in section x.x using both EW ratios and flux ratios.  Average: +1 for a positive correlation, 0 for no correlation, -1 for a negative correlation}

\end{deluxetable}

\clearpage
\begin{deluxetable}{lrrrrr}
\tabletypesize{\scriptsize}
\tablecaption{Br $\gamma$, [FeII], and H$_2$ Planefit Results}
\tablewidth{0pt}
\tablecolumns{4}
\tablehead{
\colhead{IRAS} &
\colhead{R$_0$ \tablenotemark{a}} &
\colhead{R$_1$} & 
\colhead{R$_2$} & 
}
\startdata   
03301+3111    &  $-$1.378  &    3.077  &  $-$0.746  \\
04239+2436    &     1.298  &    0.248  &  $-$0.482  \\
05256+3049    &     0.581  & $-$0.754  &     1.386  \\
16442-0930    &     2.294  & $-$0.886  &  $-$0.369  \\
21352+4307    &     1.032  & $-$1.356  &     0.919  \\

\enddata
\tablenotetext{a}{Fit to equation $\Delta$Br$\gamma$ = R$_0$ + R$_1$$\Delta$[FeII] +  R$_0$$\Delta$H$_2$, where $\Delta$ is the relative change in the line equivalent width}
\end{deluxetable}

\begin{figure}
 \plotone{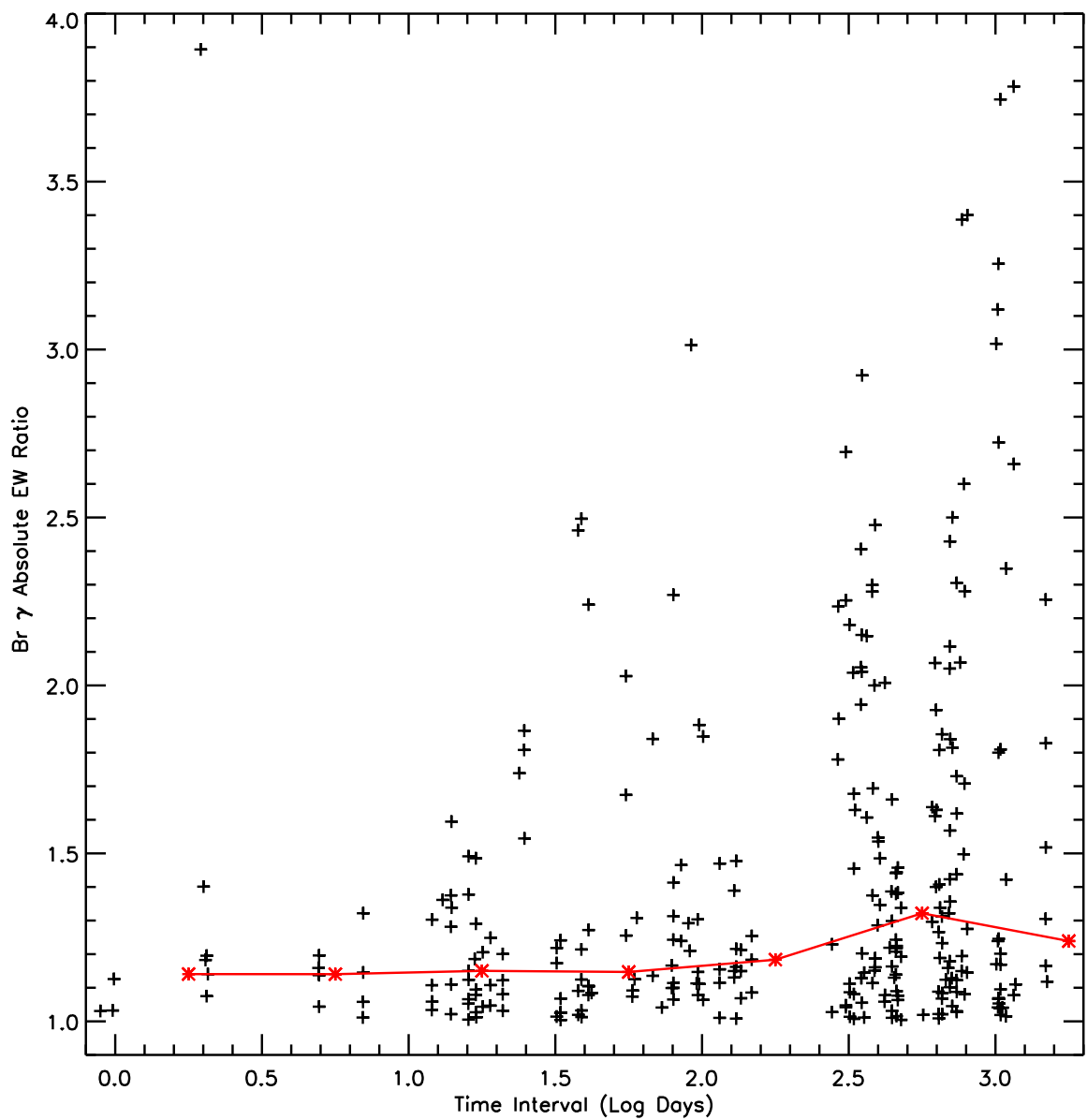}
 \caption{Absolute value of the relative variability of the EWs of Br $\gamma$ versus the time between epochs.  The solid red line is the median variability for each half-dex of time interval centered on each star symbol.  Although the maximum variability increases with time interval, the median variability does not increase significantly with time interval.  This suggests that observations over a time interval of a year or less are sampling variability with a time scale of less than a day, although larger excursions are possible.  The Br $\gamma$ EW of IRAS F23591+4748 increased by a factor of $\sim4$ within 2 days.\label{fig1}}
 \end{figure}
\clearpage

\begin{figure}
 \plotone{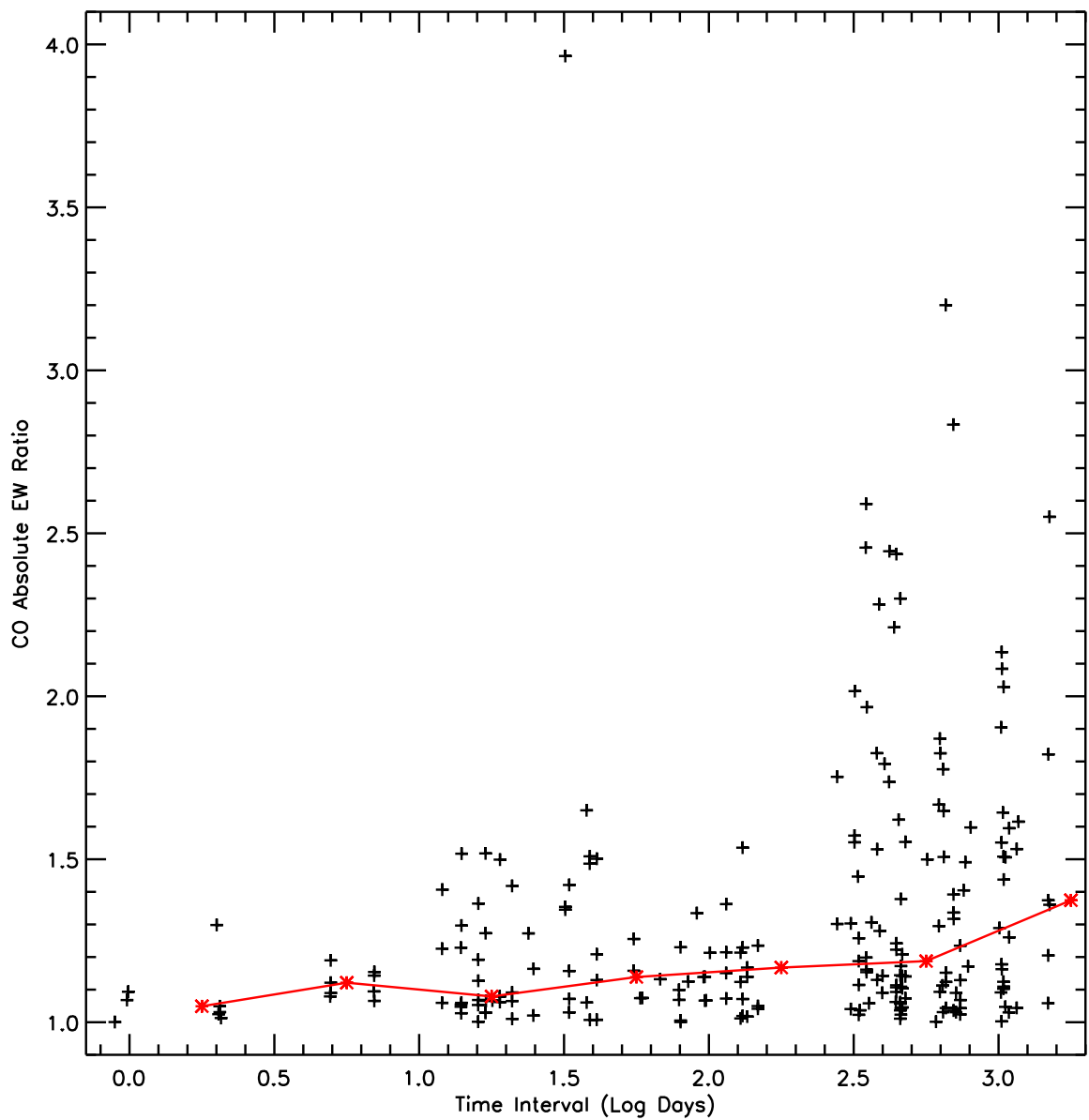}
 \caption{Absolute value of the relative variability of the EWs of CO versus the time between epochs.  The variability appears capped below $50\%$ for time scales less than one year, suggesting that observations over a time interval of a year or less are sampling variability with a time scale of roughly a day.  Variability greater than $\sim50\%$ is only seen on time scales longer than $\sim100$ days, suggesting that the CO emission originates father out into the disk than the Br $\gamma$ emission.  The CO EW of IRAS 21569+5842 increased by a factor of $\sim4$ within 30 days. \label{fig1}}
 \end{figure}
\clearpage

\begin{figure}
 \plotone{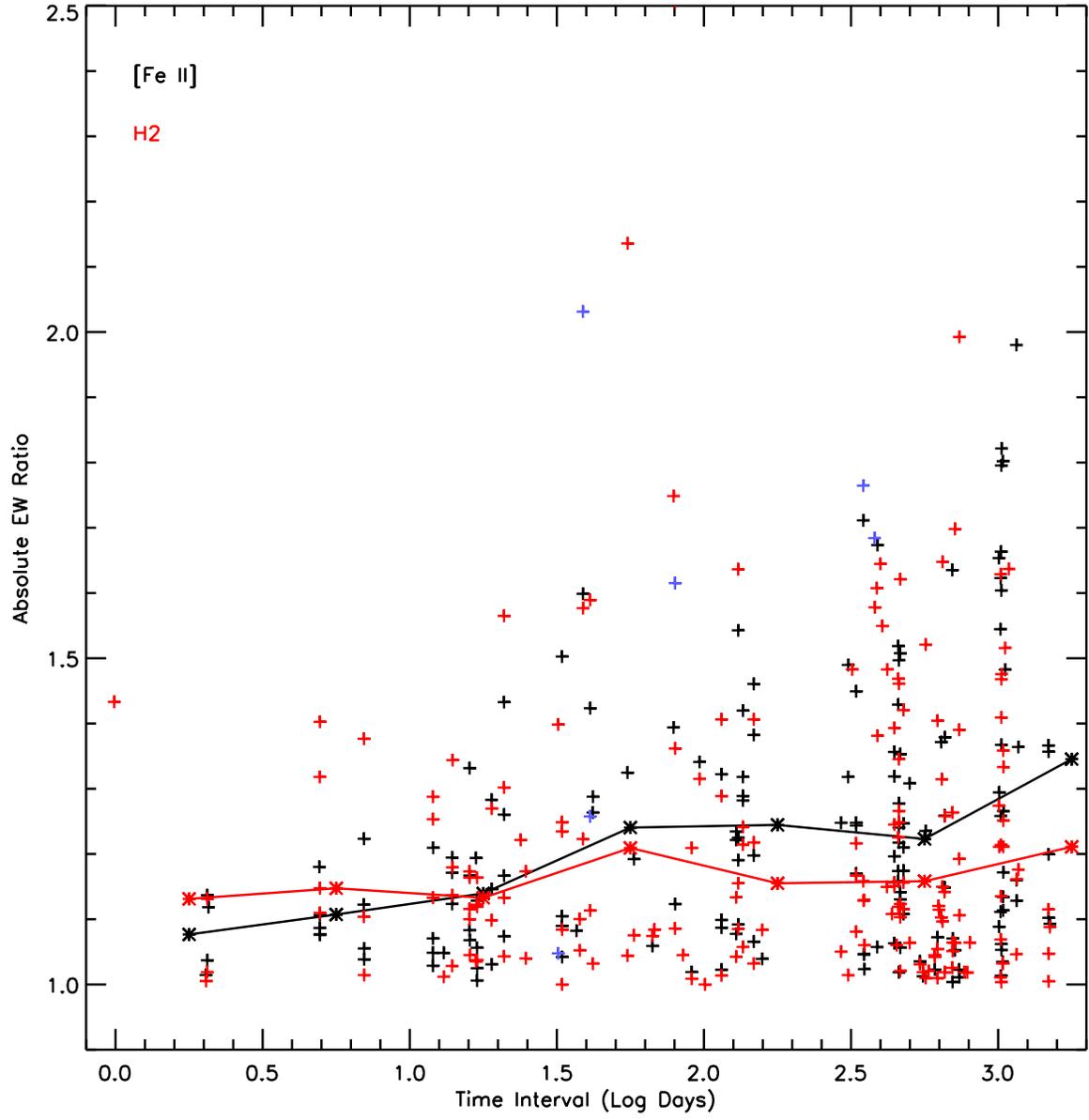}
 \caption{Absolute value of the relative variability of the EWs of [FeII] and H$_2$ versus the time between epochs. \label{fig1}}
 \end{figure}
\clearpage

\begin{figure}
  \addtocounter{figure}{-1}
\plotone{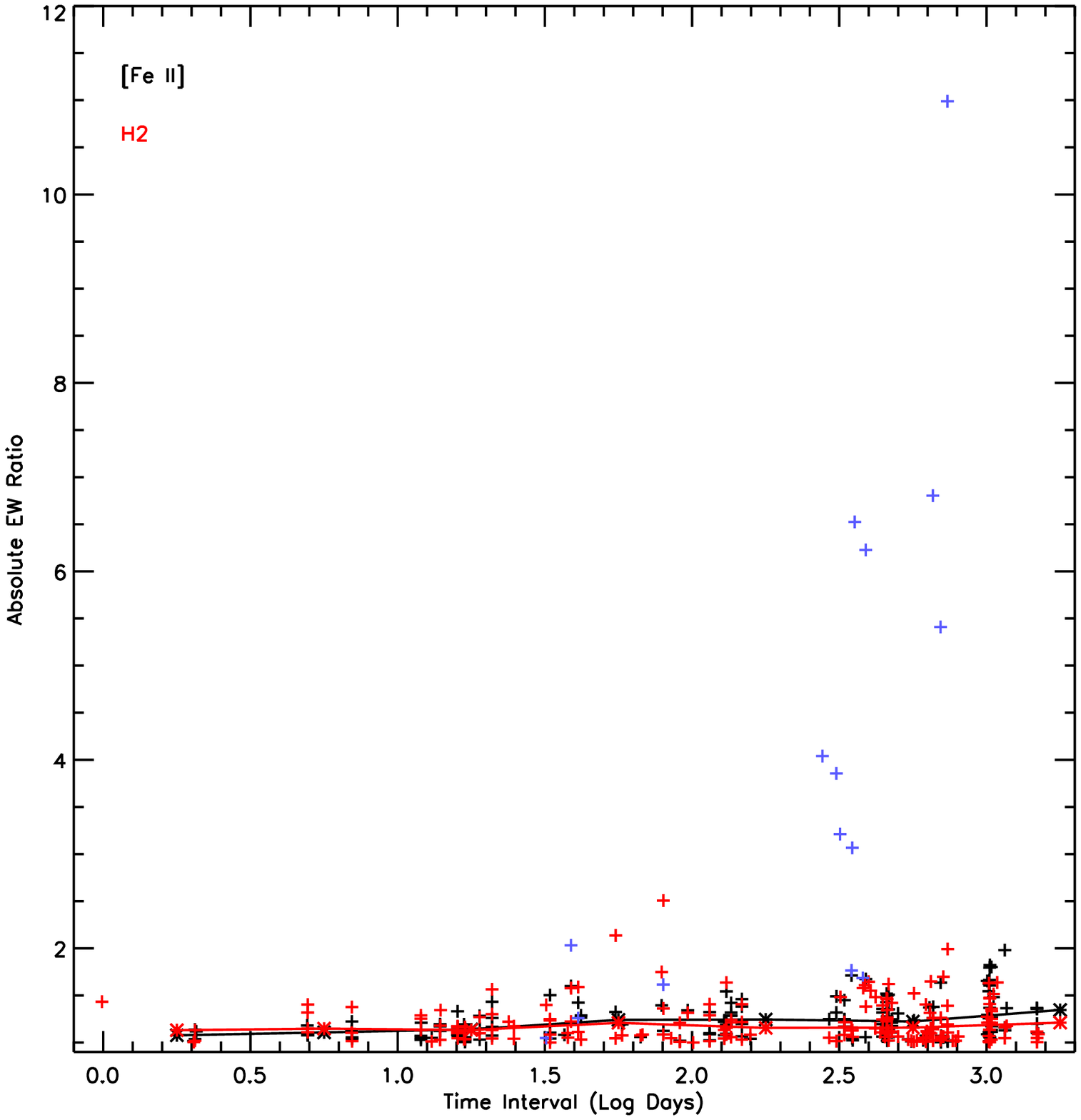}
 \caption{Absolute value of the  relative variability of the EWs of [FeII] and H$_2$ versus the time between epochs, showing the full range of the variability.  IRAS 21352+4307 (shown in blue) has much higher H$_2$ variability than any other target. \label{fig1}}
 \end{figure}
\clearpage

\begin{figure}
 \plotone{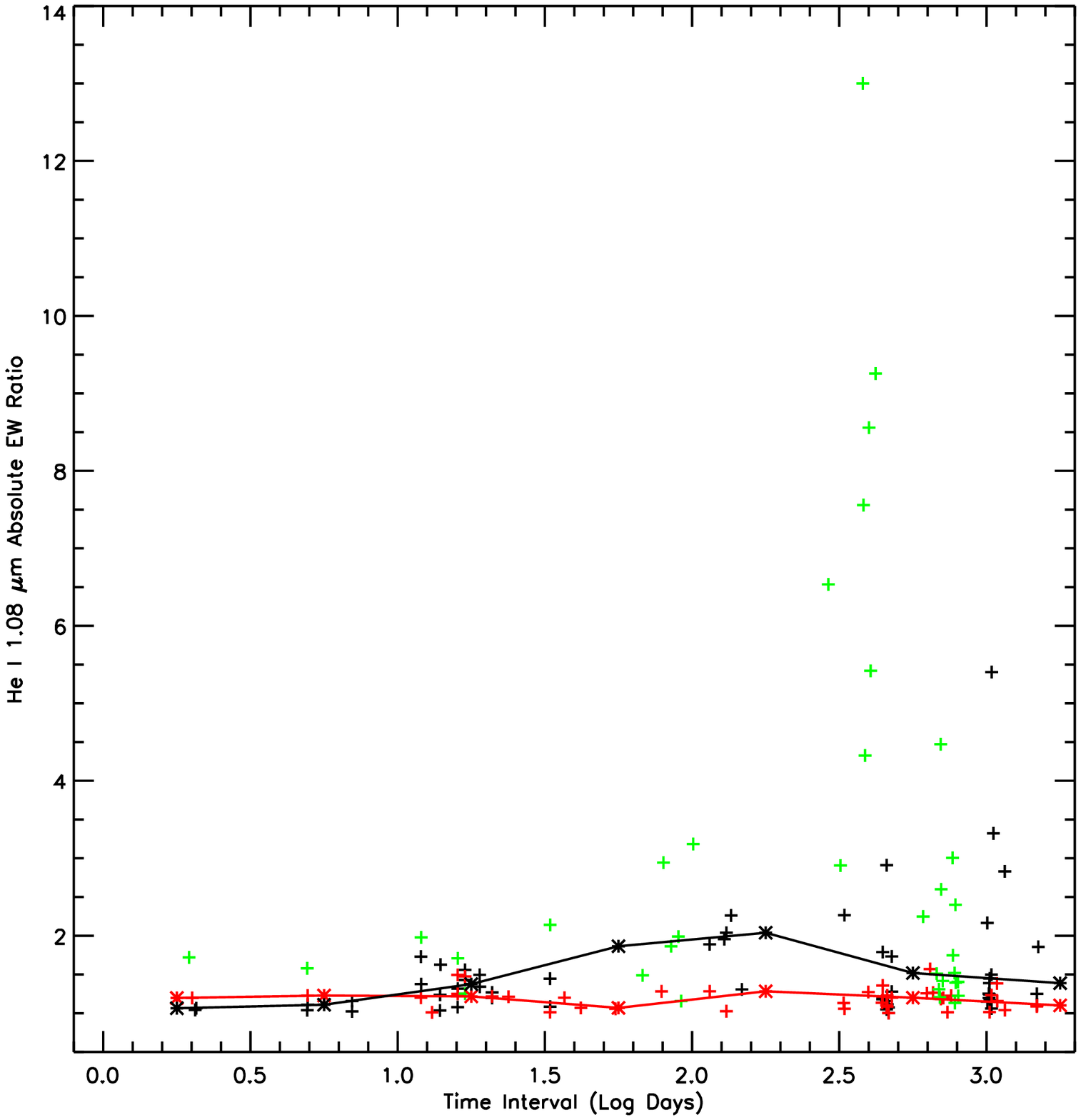}
 \caption{Absolute value of the relative variability of the EWs of He I versus the time between epochs.  The red are for targets with FU Ori-like spectra (IRAS 04287+1801, IRAS 06297+1021W, and IRAS 21454+4718), whereas the black data for the other non-FU Ori-like other targets (IRAS 03220+3035N, IRAS 03301+3111, IRAS 05513$-$1024).  The data for IRAS F23591+4748 is shown in green, which showed very high variability in the He I line.  The He I line in the stars with FU-Ori like spectra are remarkably stable.  The solid line traces the median variability for each half-dex of time interval.  \label{fig1}}
 \end{figure}
\clearpage

\begin{figure}
\plotone{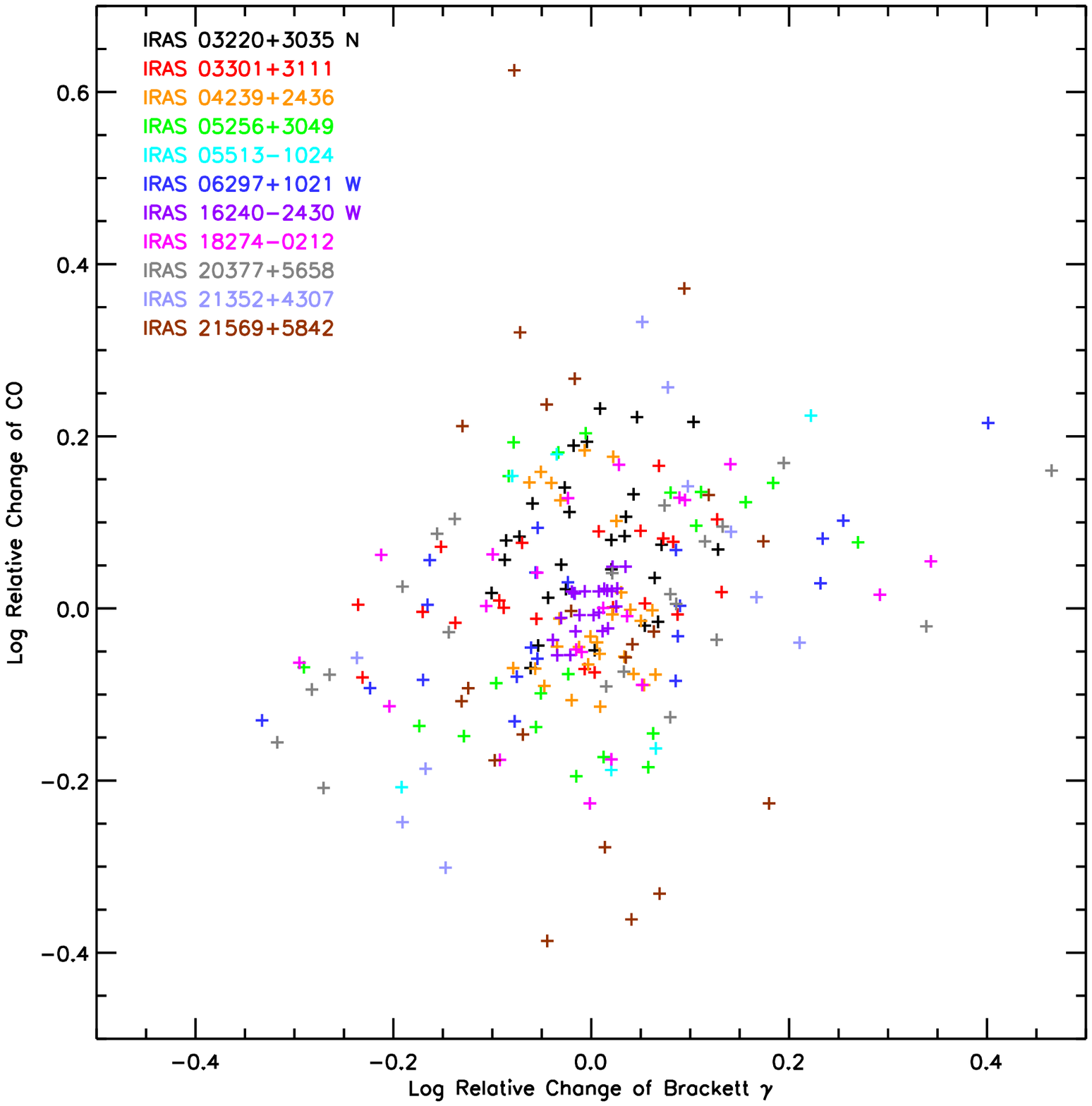}
\caption{Relative change in the EWs of Br $\gamma$ versus CO, showing no correlation in general.  Data for each target is shown in a different color.  \label{fig1}}
\end{figure}
\clearpage

\begin{figure}
\plotone{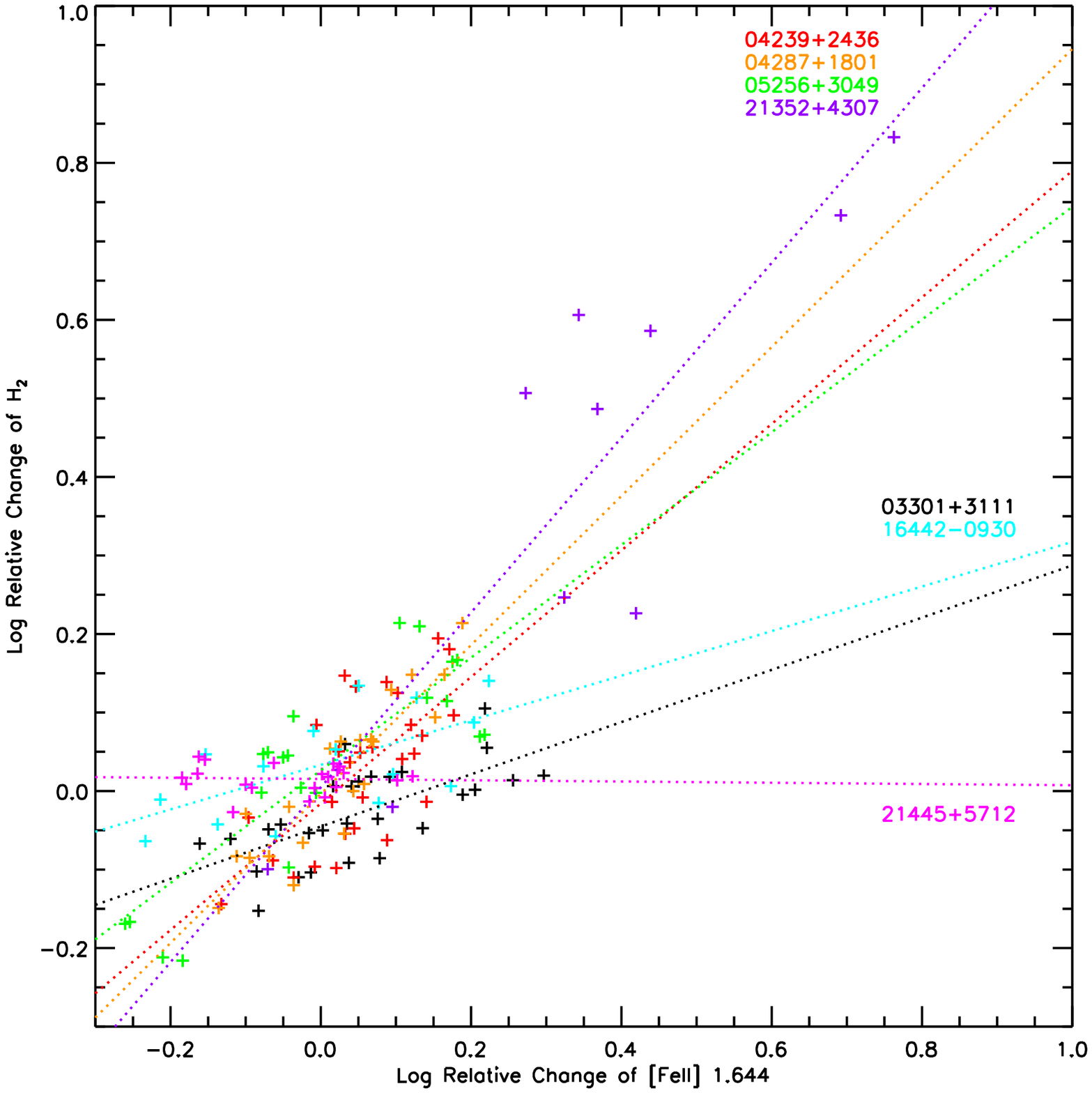}
\caption{Relative change in the EWs of [Fe II] versus H$_{2}$, suggesting a strong correlation for some targets (coded by color).  For most targets, the there is a correlation between [Fe II] versus H$_{2}$ with a best fit slope of $\sim0.8$, suggesting that the H$_{2}$ is also mostly shock excited.  For 2 targets, there is a correlation between [Fe II] versus H$_{2}$ with a best fit slope of $\sim0.3$, showing that the excitation of H$_{2}$ is dominated by a mechanism other than shock.  For one target, there is no correlation.  Error bars are omitted for clarity.  Data for each target is shown in a different color.  \label{fig1}}
 \end{figure}
\clearpage
 
\begin{figure}
\plotone{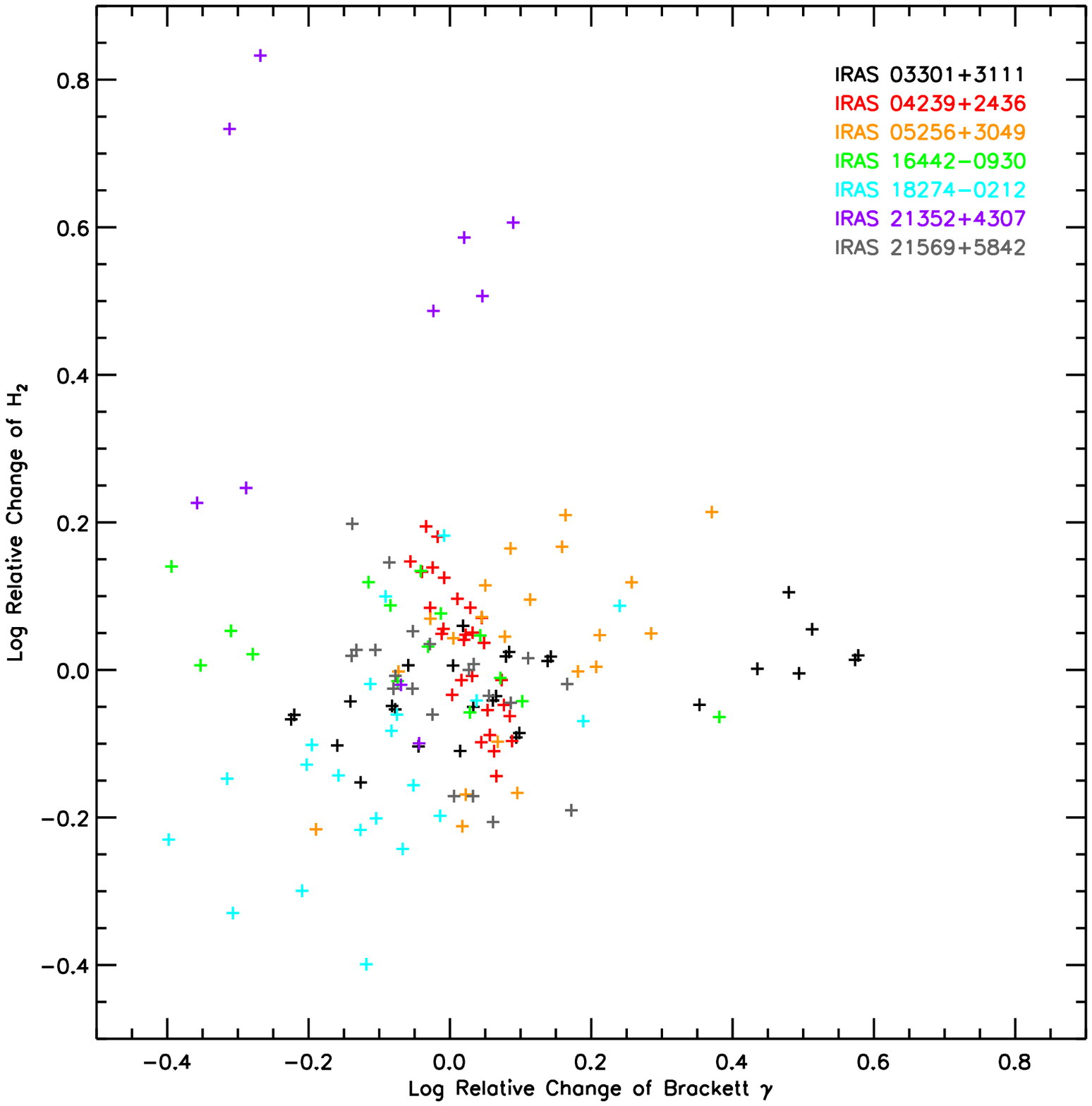}
\caption{Relative change in the EWs of Br $\gamma$ versus H$_{2}$.  Data for each target is shown in a different color.  \label{fig1}}
\end{figure}
\clearpage

\begin{figure}
\plotone{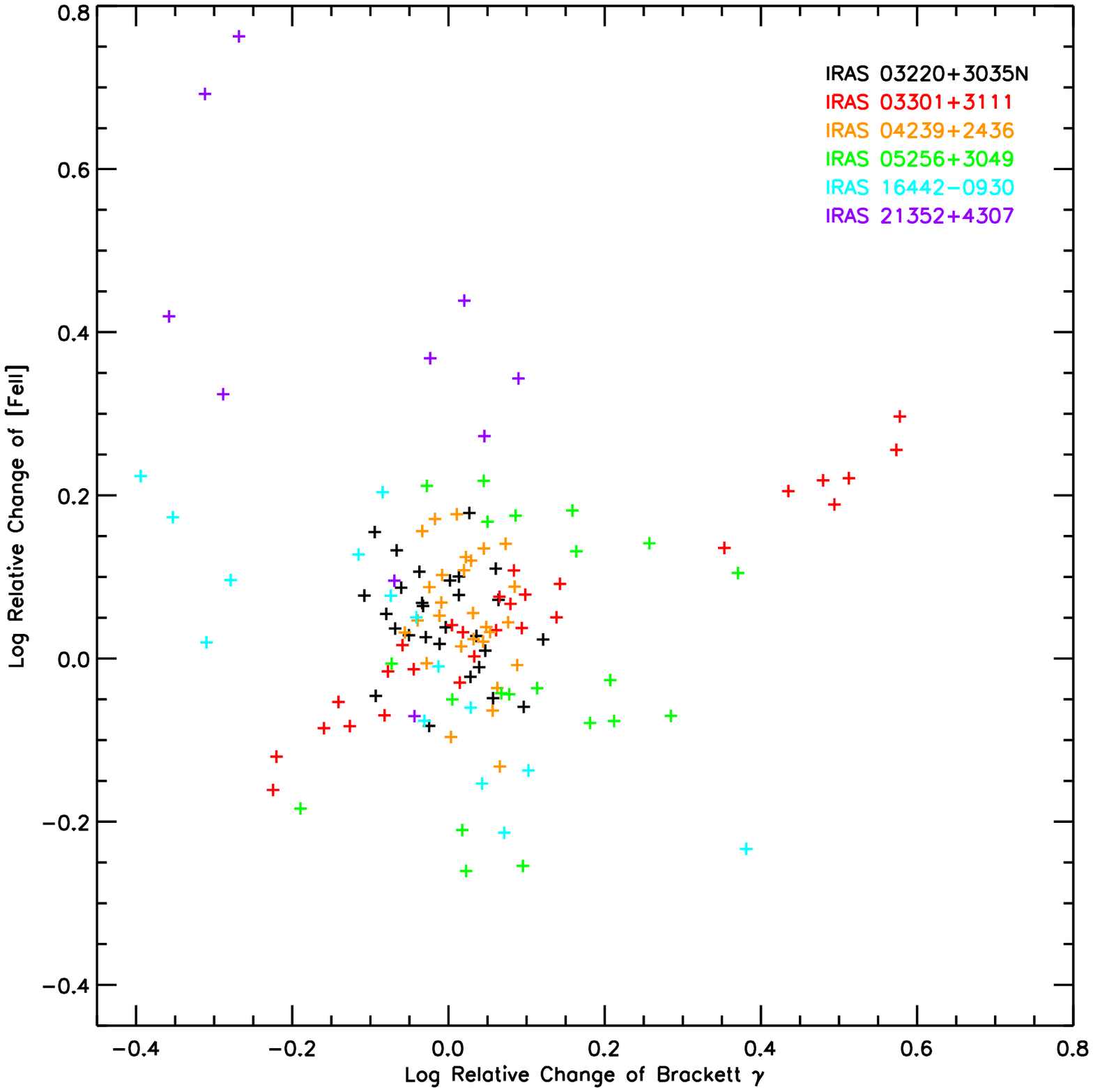}
\caption{Relative change in the EWs of Br $\gamma$ versus [FeII].  Data for each target is shown in a different color.  \label{fig1}}
\end{figure}
\clearpage

\begin{figure}
\plotone{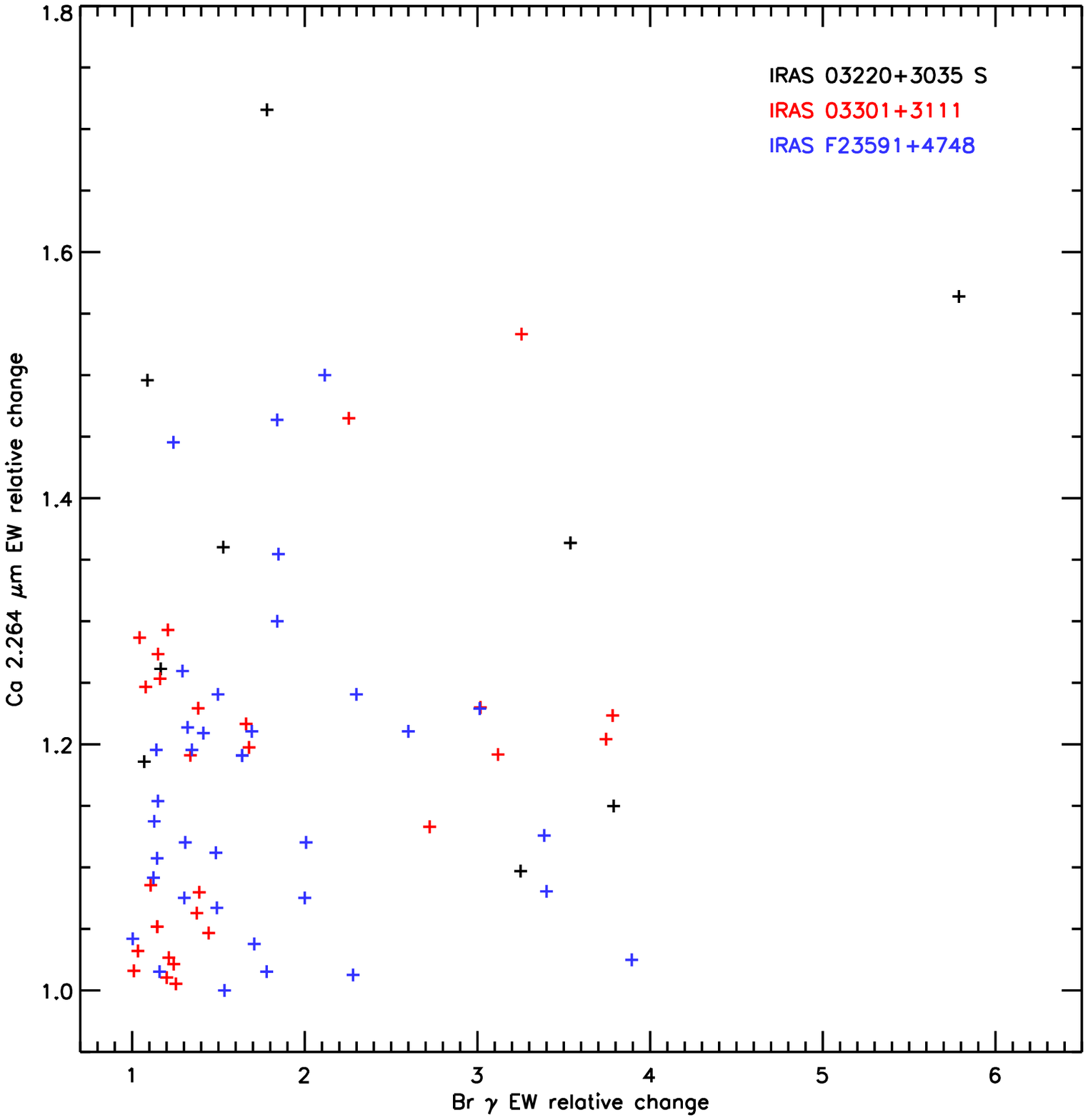}
\caption{ This plot shows the relative variability of Br $\gamma$ versus the Ca I photospheric absorption doublet at 2.264 $\mu$m, the only photospheric line that we observed to not be pushed into emission.  Each color corresponds to a different target.  If the variability of emission line EWs is simply due to changes in the continuum, then we would expect to see a correlation between the EWs of these two lines.  However, there is no correlation, suggesting that the variability in EW is due to changes in the emission line flux.   \label{fig1}}
 \end{figure}
\clearpage

\begin{figure}
 \plotone{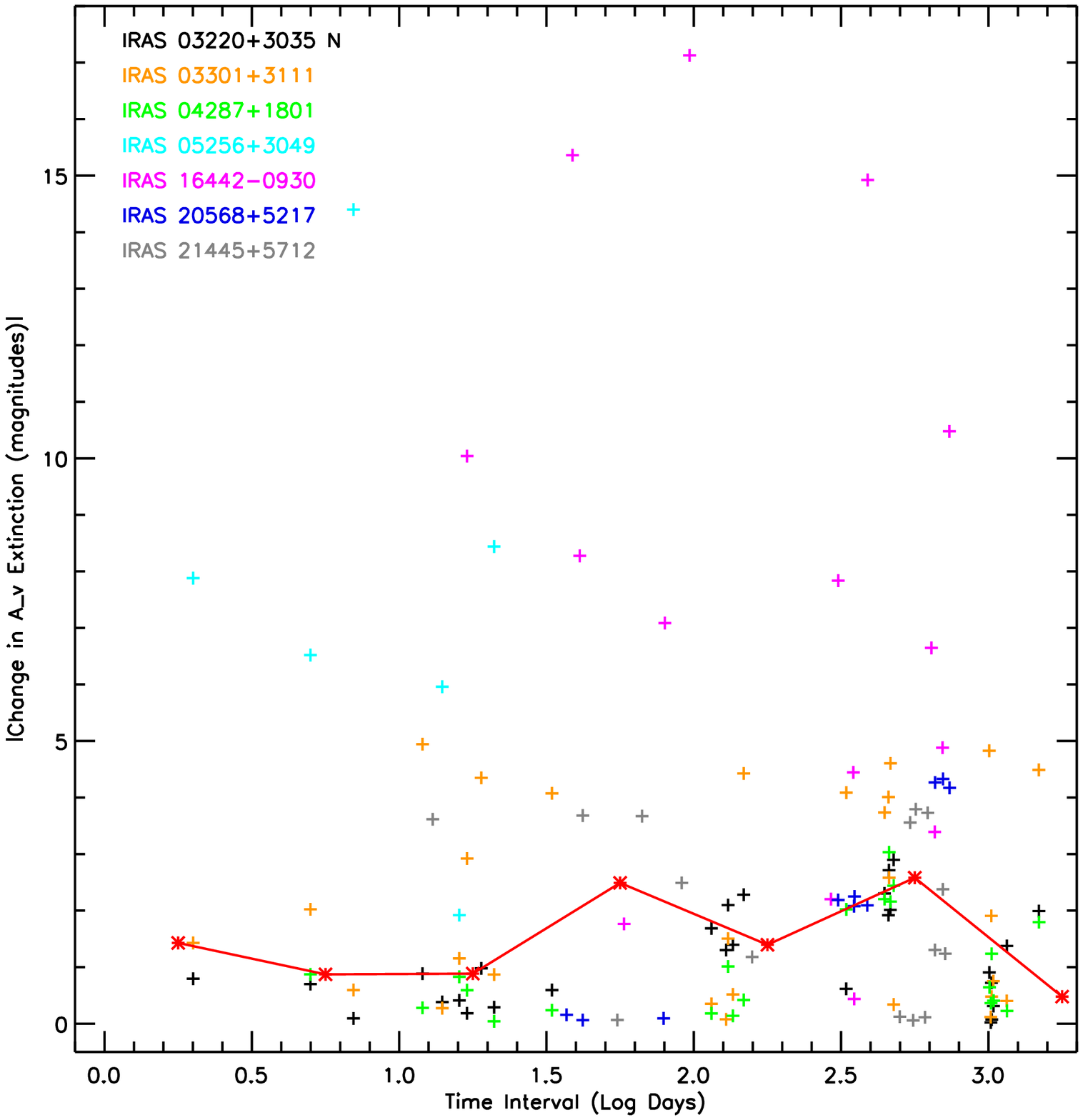}
 \caption{Absolute value of the change in the A$_v$ extinction (in magnitudes) versus the time between epochs, as traced by the ratio of the [FeII] lines at 1.256 and 1.644 $\mu$m.  Extinction can be highly variable on all time intervals, suggesting that the source of the variable extinction is close to the central star.  \label{fig1}}
 \end{figure}
\clearpage

\begin{figure}
 \plotone{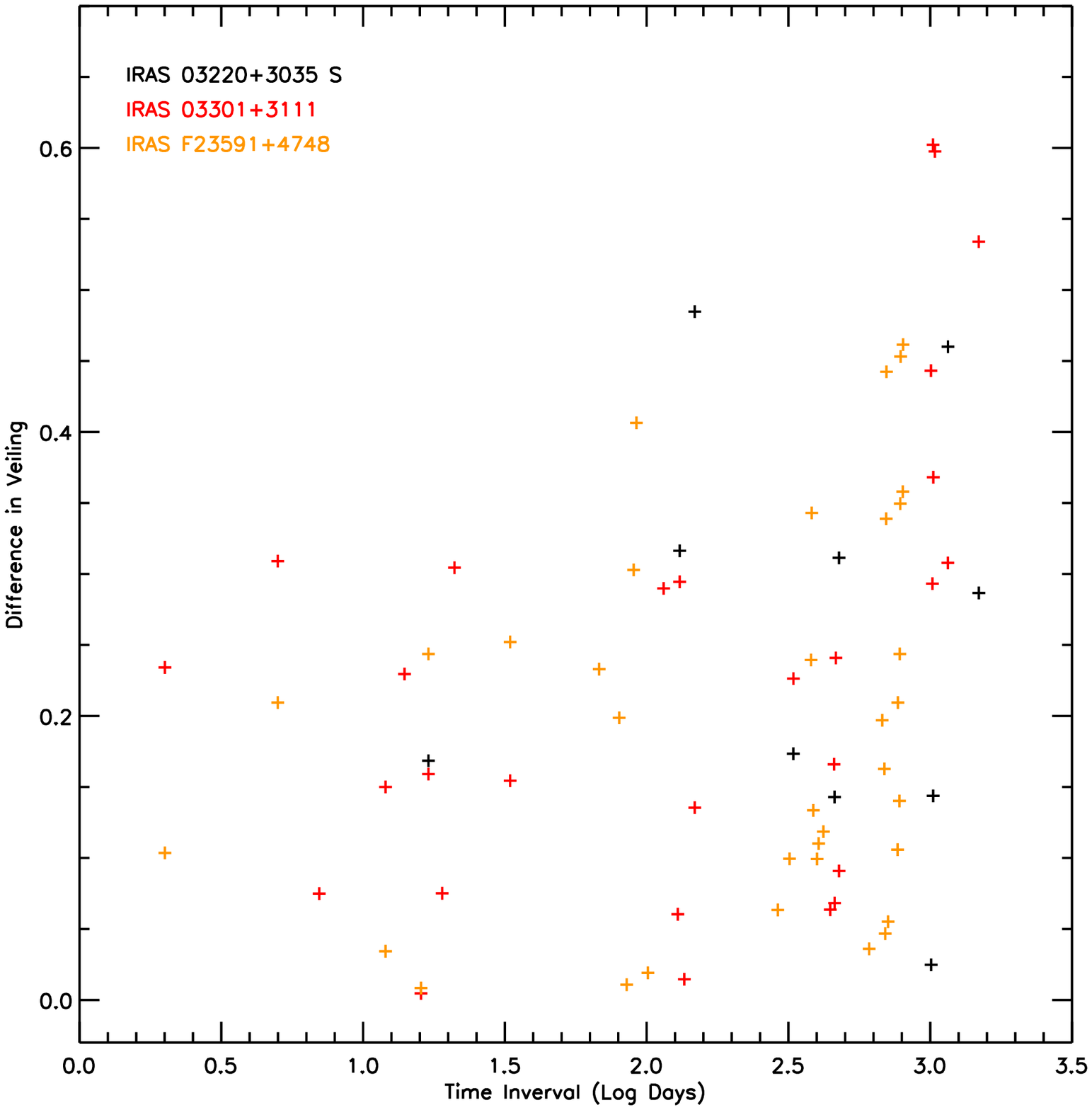}
\caption{The difference in the veiling estimate versus the time between epochs, as traced by the 1.488, 1.503, 1.711, and 2.264~$\mu$m photospheric absorption lines.  Each color represents values from a different target.  The mean difference is 0.20, with a mean uncertainty of each estimate of 0.24.  \label{fig1}}
 \end{figure}
\clearpage

 \begin{figure}
 \plotone{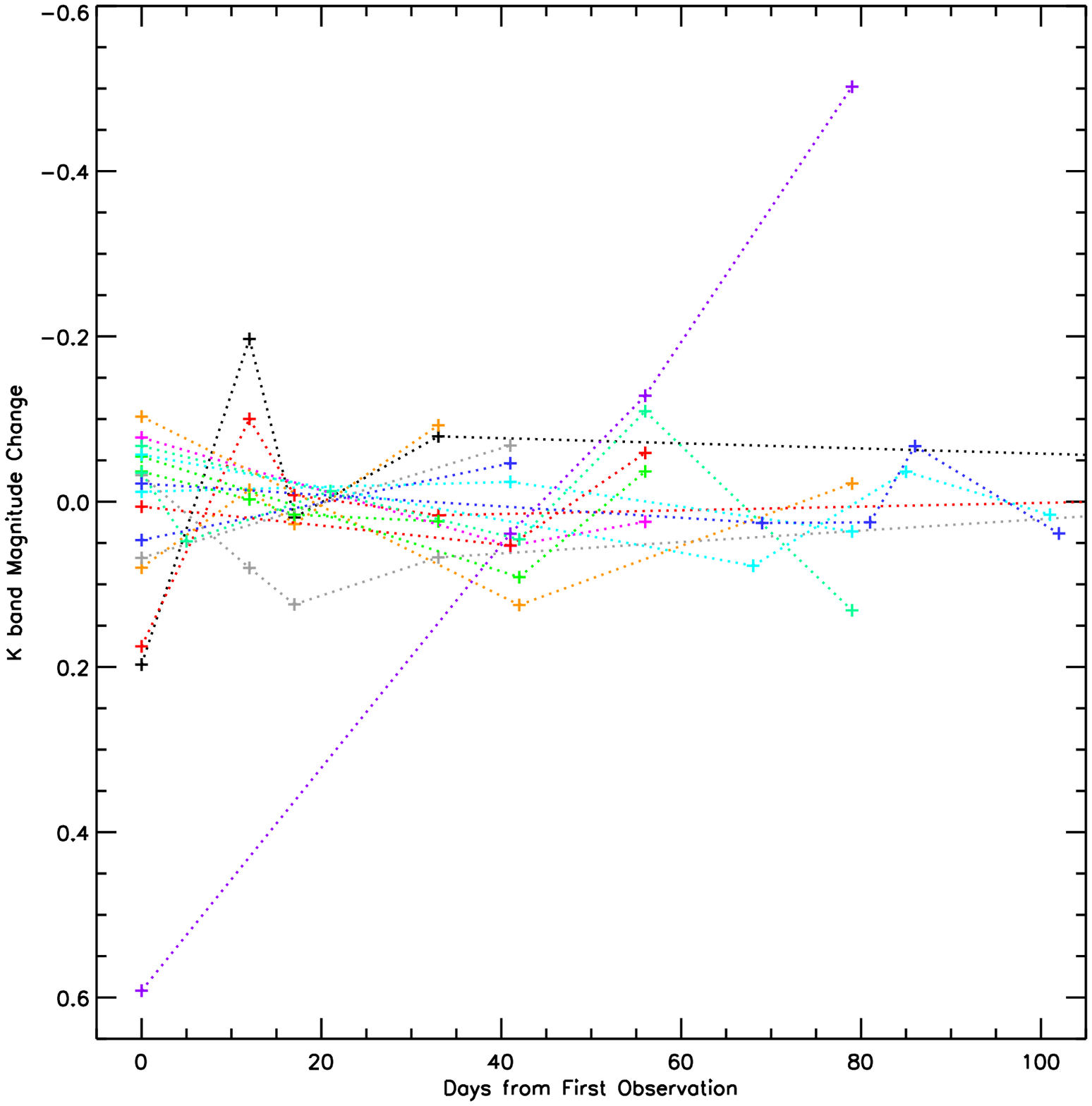}
 \caption{Plot of the K-band variability of the sample targets.  The mean photometric uncertainty is 0.055 magnitudes.  The median of the standard deviations of the K-band magnitudes is 0.080.  Each color represents a different target.  Only IRAS 18274$-$0212 showed an overall trend in its K-band photometry.  \label{fig1}}
 \end{figure}
\clearpage

\begin{figure}
 \plotone{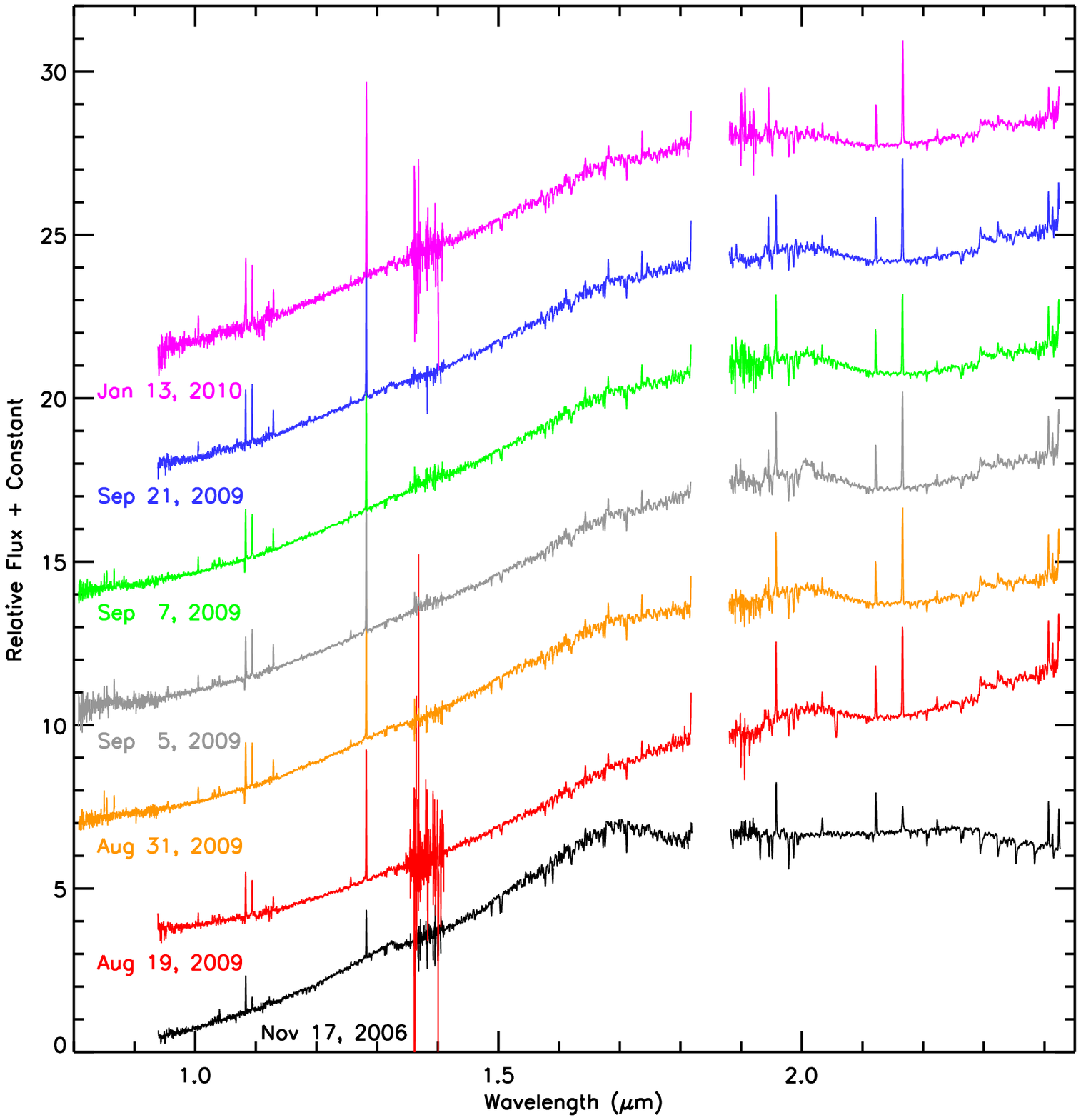}
 \caption{The spectra of IRAS 03301+3111 over 7 epochs.  This is the only target that showed a strong change in the continuum.  Between Nov 2006 and Aug 2009, the veiling increased, and CO transitioned from absorption to emission.  The fluxes were normalized to the continuum of the Nov 17, 2006 spectrum at 2.17~$\mu$m to correct for variable slit loss, then offset for clarity. \label{fig1}}
 \end{figure}
\clearpage

\begin{figure}
 \plotone{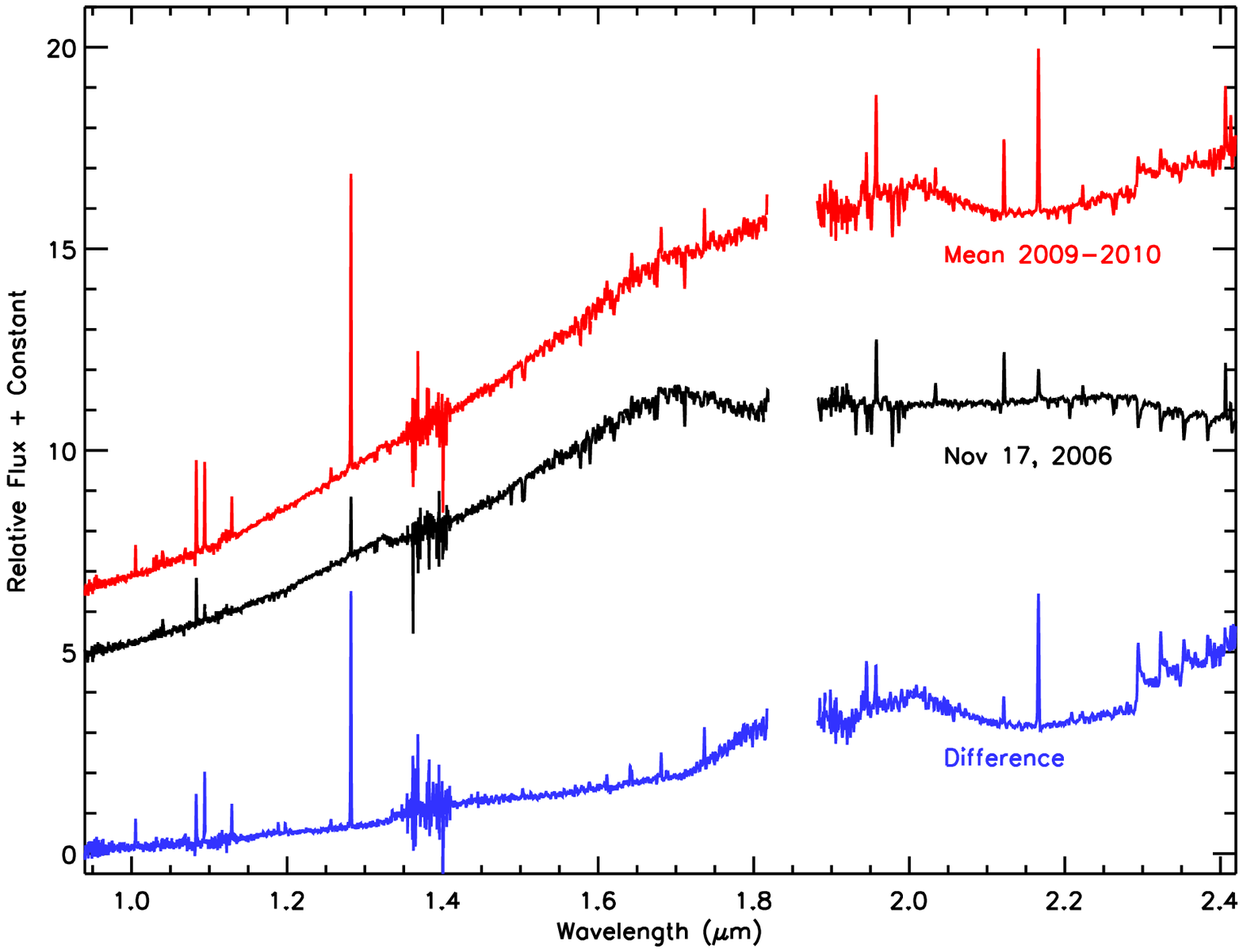}
 \caption{In this figure, we averaged together the spectra of IRAS 03301+3111 from 2009 to 2010, then subtracted the Nov. 2006 spectrum.  The averaged spectrum and 2006 spectrum both have approximately the same flux near 1~$\mu$m, and the differenced spectrum is not offset by a constant.  We can see the stronger emission lines due to H, H$_2$, CO and Ca.  We can now clearly see the increased veiling flux, and water emission from $\sim$1.7~$\mu$m to $\sim$2.1~$\mu$m.    \label{fig1}}
 \end{figure}
\clearpage

\begin{figure}
\plotone{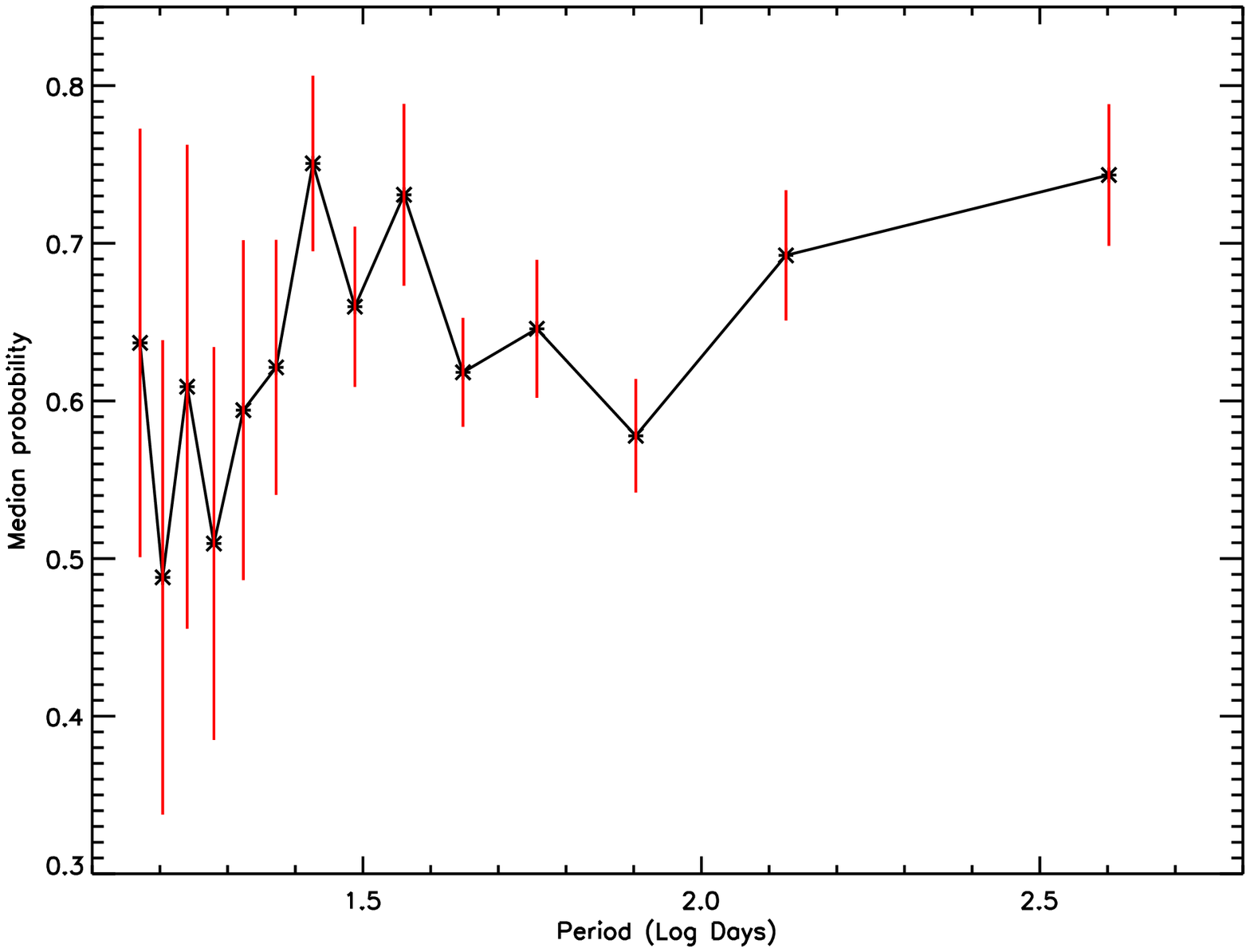}
 \caption{Periodigram of the variability of Br $\gamma$ emission.  Overall, the periodigram is nearly flat, suggesting that the variability of Br $\gamma$ emission on time scales from weeks to years is approximately white noise.  \label{fig1}}
 \end{figure}
\clearpage

\begin{figure}
 \plotone{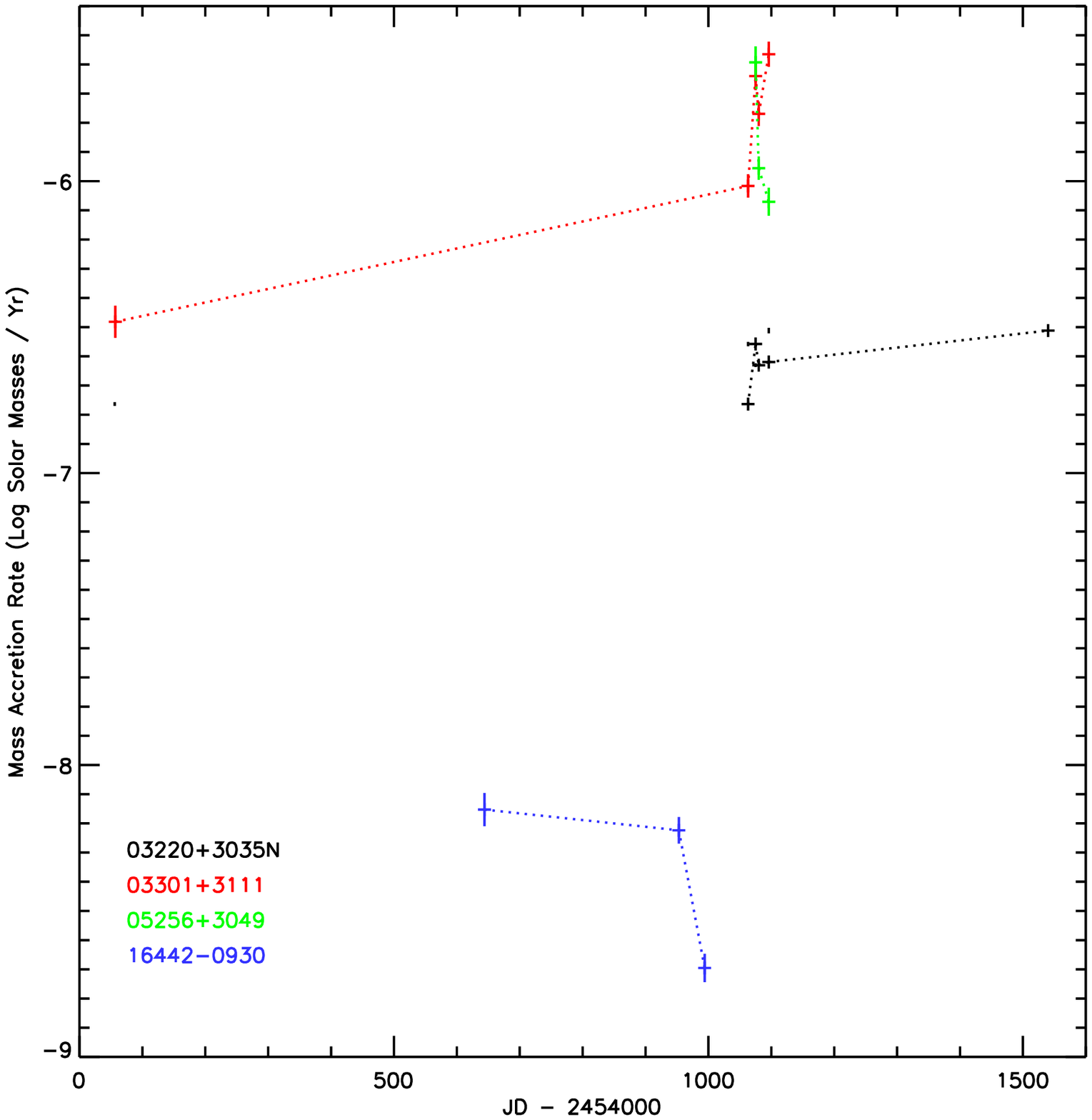}
 \caption{Plot of the mass accretion rate vs. time.  This analysis could only be done for targets exhibiting Br $\gamma$ emission as well as the 1.256~$\mu$m and 1.644~$\mu$m [FeII] lines to measure extinction, and only for nights that were photometric.  Our results show that these Class I YSOs have a wide range of mass accretion rates, and that the mass accretion rates are variable by more than a factor of 2 on a time scale of a few days.  \label{fig1}}
 \end{figure}
\clearpage

\begin{figure}
 \plotone{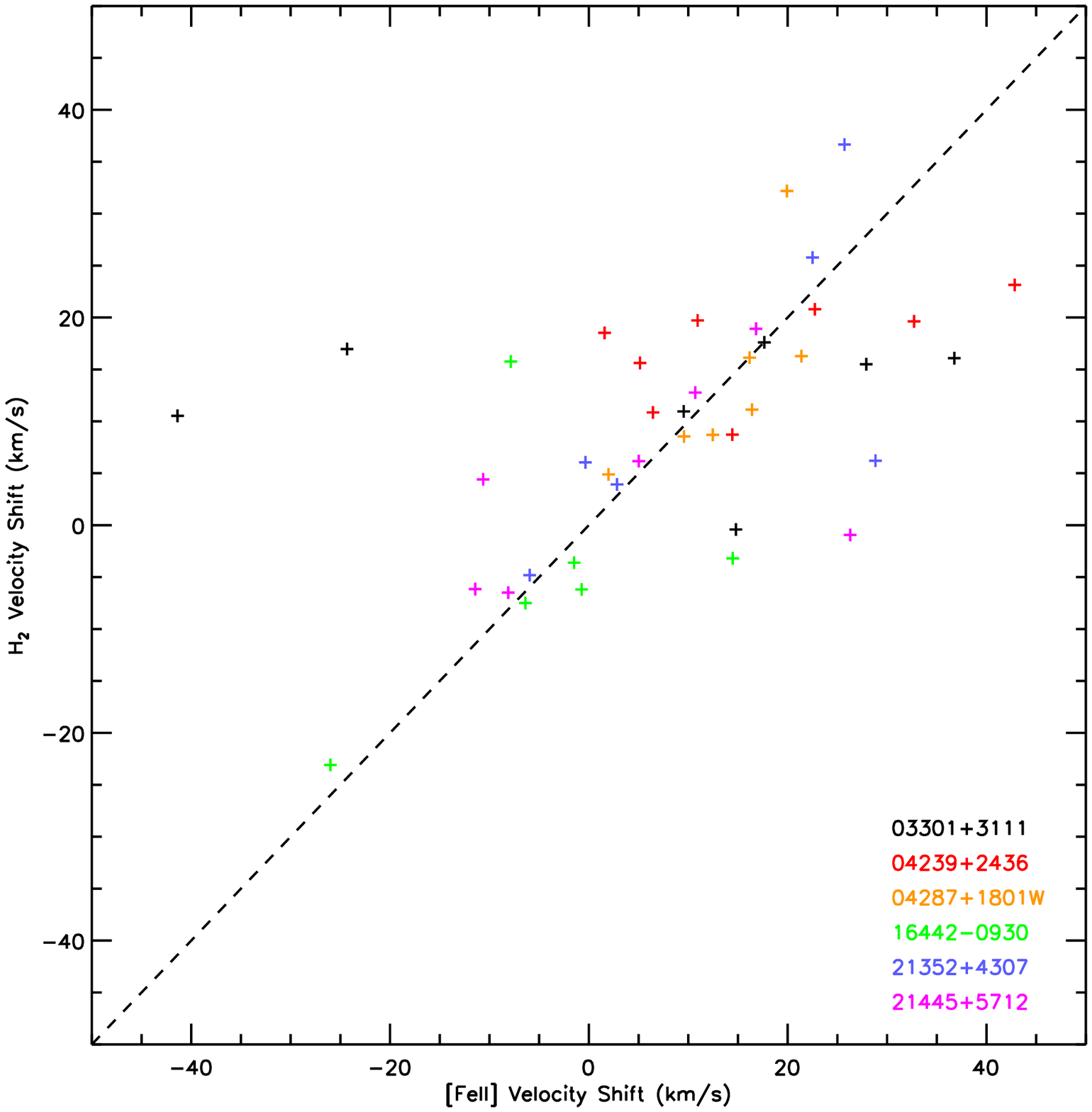}
 \caption{The velocity shift in the [FeII] line versus the H$_2$ line.  There is generally a good correlation between the velocity shifts of these two lines, supporting the conclusion that they have the same excitation source.  The mean uncertainty in the velocity is X kms$^{-1}$.  IRAS 03301+3111 is the only target with both [FeII] and H$_2$ emission where the correlation coefficient less than 0.5.  The dashed line has a slope of 1 and passes through (0,0), and is the expected regression line if the radial velocities of these two lines vary in step with each other.  \label{fig1}}
 \end{figure}
\clearpage

 \begin{figure}
 \plotone{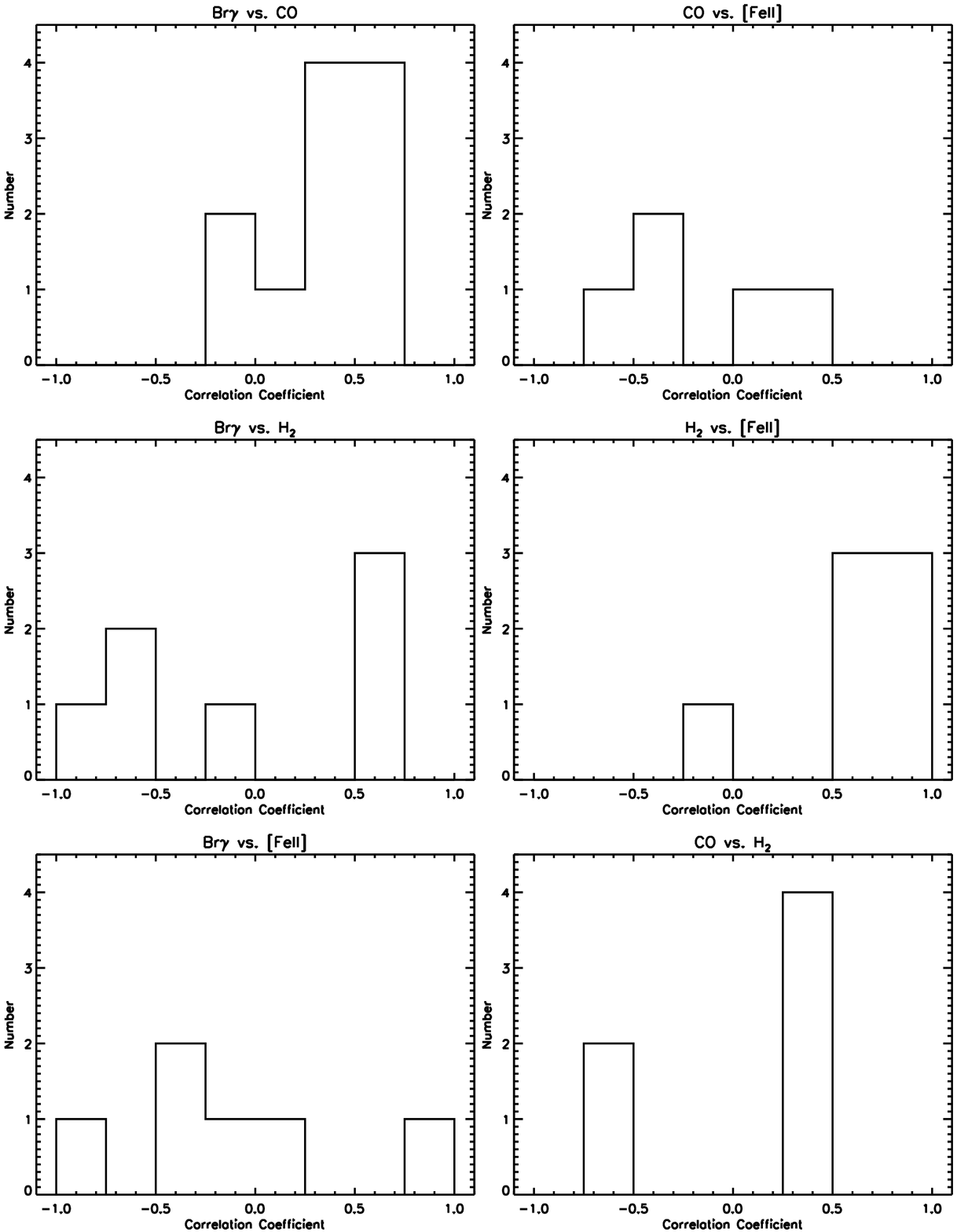}
 \caption{Histograms of correlation coefficients between the relative change of mass accretion and wind tracers.  H$_2$ vs. [FeII] mostly has strong positive correlations.  \label{fig1}}
 \end{figure}
\clearpage

\begin{figure}
 \plotone{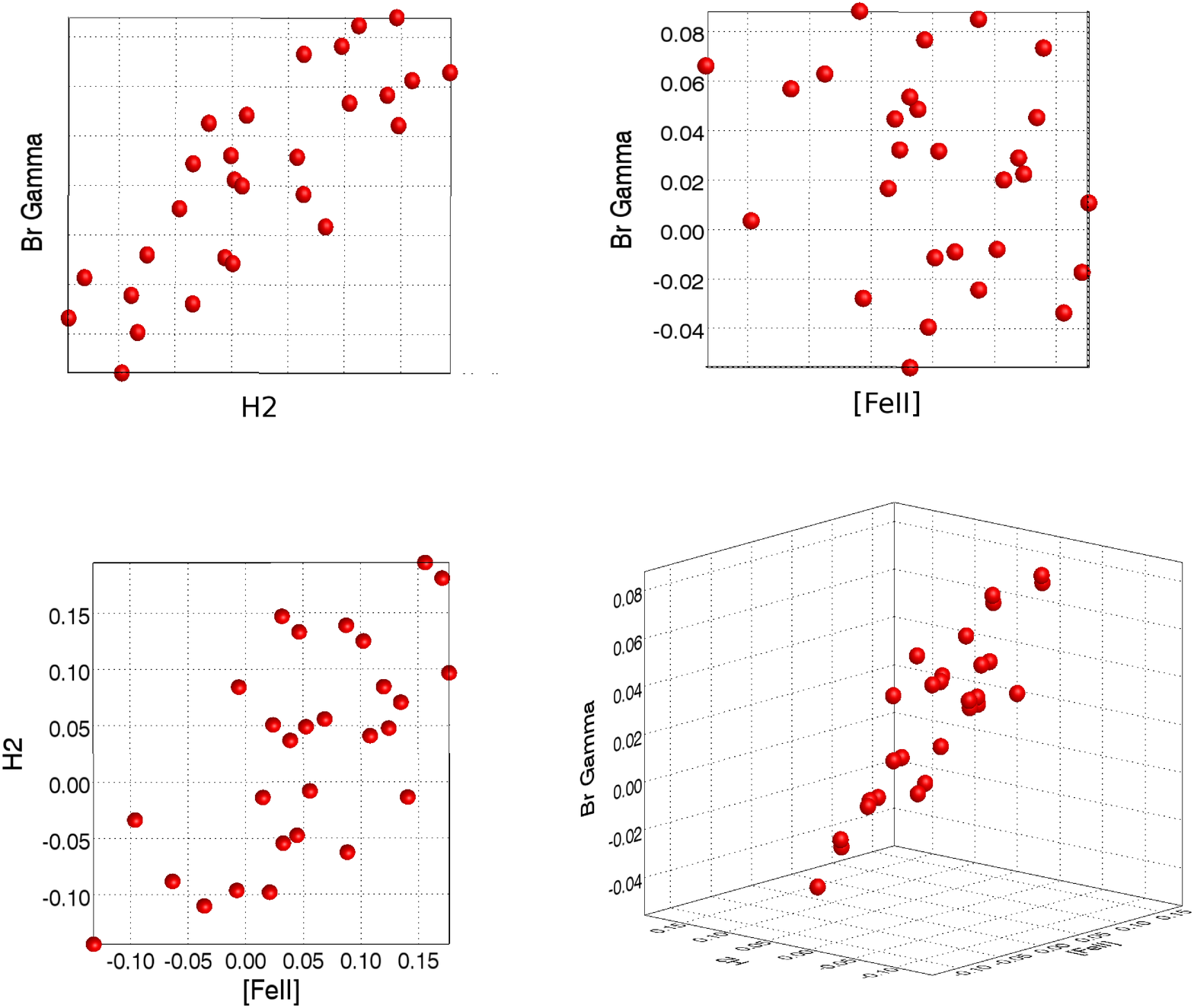}
 \caption{Example of a 3D plot of Br $\gamma$ vs. H$_2$ vs. [FeII] of IRAS 04239+2436, showing how the data lie on a plane in 3D.  The top left panel shows the  Br $\gamma$ vs. H$_2$ projection, the top right panel shows the Br $\gamma$ vs. [FeII] projection, the bottom left panel shows the H$_2$ vs. [FeII] projection, and the lower right panel shows the space rotated to minimize the projected scatter of the data points.  \label{fig1}}
 \end{figure}
\clearpage


\clearpage

\begin{thebibliography}{}
\bibitem[()]{} 

\bibitem[Alves de Oliveira \& Casali(2008)]{Alv2008} Alves de Oliveira, C., \& Casali, M., A\&A, 485, 155

\bibitem[Artemenko et al.(2010)]{Art2010} Artemenko, S., Grankin, K., \& Petrov, P., 2010, Astronomy Reports, 54, 163

\bibitem[Bally et al.(2007)]{Bal2007} Bally, J., Reipurth, B., \& Davis, C., in Protostars and Planets V, eds. B. Reipurth, D. Jewitt, \& K. Keil, University of Arizona Press, 2007, p.215

\bibitem[Barsony et al.(2005)]{Bar2005} Barsony, M., Ressler, M., Marsh, \& Kenneth A., 2005, ApJ, 630, 381

\bibitem[Bary et al.(2009)]{Bar2009} Bary, J., Leisenring, J., \& Skrutskie, M., 2009, ApJ, 706, L168

\bibitem[Beck et al.(2008)]{Beck2008} Beck, T., McGregor, P., Takami, M., \& Pyo, T., 2008, ApJ, 676, 472 

\bibitem[Biscaya et al.(1997)]{Bis1997} Biscaya, A., Reike, G., Narayanan, G., Luhman, K., \& Young, E., 1997, ApJ, 491, 359

\bibitem[Brittain et al.(2010)]{Bri2010} Brittain, S., Rettig, T., Simon, T., Gibb, E., \& Liskowsky, J., 2010, ApJ, 708, 109

\bibitem[Calvet et al.(1991)]{Cal1991} Calvet, N., Pati\~{n}o, A., Magris, G., \& D'Alessio, P., 1991, ApJ, 380, 617 

\bibitem[Carpenter et al.(2001)]{Car2001} Carpenter, J., Hillenbrand, L., \& Skrutskie, M., 2001, AJ, 121, 3160

\bibitem[Choudhury et al.(2011)]{Cho2011} Choudhury, R., Bhatt, H., \& Pandey, G., 2011, A\&A, 526, 97

\bibitem[Connelley et al.(2007)]{Con2007} Connelley, M., Reipurth, B., \& Tokunaga, A., 2007, \aj, 133, 1528 




\bibitem[Connelley \& Greene(2010)]{Con2010} Connelley, M., \& Greene, T., 2010, \aj, 140 1214 

\bibitem[Cushing et al.(2004)]{Cus2004} Cushing, M., Vacca, W., \& Rayner, J., 2004, \pasp, 116, 362  

\bibitem[Davis et al.(2001)]{Dav2001} Davis, C., Ray, T., Desroches, L., \& Aspin, C., 2001, MNRAS, 326, 524

\bibitem[Dullemond et al.(2007)]{Dul2007}Dullemond, C., Hollenbach D., Kamp, I., \& D'Alessio, P., in Protostars and Planets V, eds. B. Reipurth, D. Jewitt, \& K. Keil, University of Arizona Press, 2007, p.555

\bibitem[Edwards et al.(2006)]{Edw2006} Edwards, S., Fischer, W., Hillenbrand, L., \& Kwan, J., 2006, ApJ, 646, 319

\bibitem[Edwards et al.(2003)]{Edw2003} Edwards, S., Fischer, W., Kwan, J. Hillenbrand, L., \& Dupree, A., 2003, ApJ, 599, L41

\bibitem[Eiroa et al.(2002)]{Eir2002} Eiroa, C., et al., 2002, A\&A, 384, 1038

\bibitem[Faesi et al.(2012)]{Fae2012} Faesi, C., et al., 2012, PASP, 124, 1137   

\bibitem[Grankin et al.(2007)]{Gra2007} Grankin K., Melinkov, S., Bouvier, J., Herbst, W., \& Shevchenko, V., 2007, A\&A, 461, 183

\bibitem[Greene et al.(2010)]{Gre2010} Greene, T., Barsony, M., \& Weintraub, D., 2010, ApJ, 725, 1100

\bibitem[Greene \& Lada(2002)]{Gre2002} Greene, T., \& Lada, C., 2002, AJ, 124, 2185

\bibitem[Herbig(2002)]{Her2002} Herbig, G., in Physics of star formation in galaxies. Saas-Fee Advanced Course 29, Berlin: Springer, 2002

\bibitem[Joy(1945)]{Joy1945} Joy, A., 1945, ApJ, 102, 168

\bibitem[Kwan et al.(2007)]{Kwa2007} Kwan, J., Edwards, S., \& Fischer, W., 2007, ApJ, 657, 897

\bibitem[Manset et al.(2009)]{Man2009} Manset, N., Bastien, P., M\'{e}nard, F., Bertout, C., Le van Suum A., \& Boivin, L., 2009, A\&A, 499, 137

\bibitem[Martin(1997)]{Mar1997} Martin, S., 1997, ApJ, 478, L33 

\bibitem[Mathis(2000)]{Mat2000} Mathis, J., 2000, p. in Allen's Astrophysical Quantities, ed. Arthur Cox, The Athlone Press, London, 2000, p.523

\bibitem[Megeath et al.(2013)]{Meg2013} Megeath, S., et al., 2013, AJ, in press

\bibitem[Morales-Calder\'{o}n et al.(2011)]{Mor2011} Morales-Calder\'{o}n et al., 2011, ApJ, 733, 50   

\bibitem[Muzerolle et al.(1998)]{Muz1998} Muzzerolle, J., Hartmann, L., \& Calvet, N., 1998, AJ, 116, 2965

\bibitem[Muzerolle et al.(2009)]{Muz2009} Muzzerolle, J., et al., 2009, ApJ, 704, L15

\bibitem[Najita et al.(1996)]{Naj1996} Najita, J., Carr, J., Glassgold, A., Shu, F., \& Tokunaga, A., 1996, \apj, 462, 919


\bibitem[Nguyen et al.(2009)]{Ngu2009} Nguyen, D., Scholtz, A., van Kerkwijk, M., Jayawardhana, R., \& Brandeker, A., 2009, ApJ, 694, L153

\bibitem[Rayner et al.(2003)]{Ray2003} Rayner, J., Toomey, D., Onaka, P., Denault, A., Stahlberger, W., Vacca, W., Cushing, M., \& Wang, S., 2003, \pasp, 15, 362

\bibitem[Scholz(2012)]{Sch2012} Scholz, A., 2012, MNRAS, 420, 1495

\bibitem[Simons \& Tokunaga(2002)]{Sim2002} Simons, D., \& Tokunaga, A., 2002, PASP, 114, 169

\bibitem[Tokunaga \& Simons(2002)]{Tok2002} Tokunaga, A., \& Simons, D., 2002, PASP, 114, 180

\bibitem[Wall \& Jenkins(2008)]{Wal2008} Wall, J., \& Jenkins, C., \textit{Practical Statistics for Astronomers}, New York, Cambridge University Press, 2008 

\bibitem[Zinnecker et al.(1998)]{Zin1998} Zinnecker, H., McCaughrean, M., \& Rayner, J., 1998, Nature, 394, 862

\end{thebibliography}
\end{document}